\documentclass[%
 preprint,
 superscriptaddress,
 amsmath,amssymb,
 aps,
 showkeys,
 titlepage,
 endfloats*.   
]{revtex4-2}

\usepackage[utf8]{inputenc}
\usepackage[english]{babel}
\usepackage{multirow}

\usepackage{graphicx}
\usepackage{dcolumn}
\usepackage{bm}

\usepackage[colorlinks = true,
            linkcolor = blue,
            urlcolor  = blue,
            citecolor = red,
            anchorcolor = blue]{hyperref}

\begin{document}


\title{Impact of Phonon-Magnon Scattering on Thermal Transport across Antiferromagnetic and Ferromagnetic Oxides} 

\author{Zeyu Xiang}
\affiliation{Department of Mechanical Engineering, University of California, Santa Barbara, CA 93106, USA}

\author{Bolin Liao}
\email{bliao@ucsb.edu} \affiliation{Department of Mechanical Engineering, University of California, Santa Barbara, CA 93106, USA}

\date{\today}

\begin{abstract}
Phonon--magnon coupling has been extensively studied in ferromagnets, whereas its microscopic mechanisms in antiferromagnets remain less understood.
Here, we combine first-principles calculations with spin--lattice dynamics to investigate phonon--magnon interactions and their impact on thermal transport in antiferromagnetic NiO and MnO, which are put in contrast to the ferromagnetic EuO with an identical crystal structure.
Green--Kubo and spectral energy density analyses show that spin dynamics reduce phonon thermal conductivity and lifetimes, with the strongest effects in MnO.
Comparison of the excitation spectra indicates that spin-induced phonon scattering depends sensitively on the relative phonon and magnon energy scales, while the contrasting behavior of MnO and EuO further highlights the role of magnetic order.
A simplified phase-space model further illustrates that, for comparable phonon and magnon energy scales, the antiferromagnetic case can support a larger phonon--magnon scattering phase space than the ferromagnetic case, suggesting a greater number of kinematically allowed scattering channels.
These results identify excitation energy scales and magnetic order as key factors governing phonon--magnon scattering in magnetic materials.
\end{abstract}

\keywords{Spin-lattice Coupling, Spin-lattice Dynamics, Antiferromagnets}
                            
\maketitle


\section{Introduction}
Magnons, the quanta of spin-wave excitations, represent collective fluctuations of magnetic order and can occupy frequency ranges and exhibit characteristic timescales comparable to those of phonons, enabling potentially strong dynamical coupling between spin and lattice degrees of freedom. 
The role of spin–phonon coupling in thermal transport has therefore attracted increasing attention because of the broad implications for spin caloritronics~\cite{boona2014spin}, magnonics~\cite{chumak2015magnon,kajiwara2010transmission}, and thermal management in magnetic materials and devices~\cite{sanders1977effect,boona2014magnon,liao2014generalized}.
Experimentally, anomalies in thermal conductivity have been observed across magnetic ordering transitions in both antiferromagnetic and ferromagnetic materials.
For example, upon cooling through the N\'eel temperature, the thermal conductivity of antiferromagnetic YMnO$_3$ increases abruptly~\cite{sharma2004thermal}, whereas FeCl$_2$ exhibits a pronounced nonmonotonic temperature dependence associated with resonant magnon--phonon interactions~\cite{laurence1973thermal}.
In ferromagnetic Fe, the thermal conductivity similarly exhibits a sharp change near the Curie temperature~\cite{backlund1961experimental}.
Comparable changes in thermal transport can also be induced by magnetic fields.
Field dependence of the thermal conductivity has been reported in magnetic materials such as K$_2$V$_3$O$_8$, EuS, Gd, and $\alpha$-RuCl$_3$, where magnetic-field-induced changes in the magnetic state and excitation spectrum modify the coupling and scattering between magnetic excitations and phonons, and can thereby substantially alter thermal transport~\cite{sales2002magnetic,mccollum1964spin,leahy2017anomalous,zhang2024room}.
Together, these observations demonstrate that spin--phonon coupling can substantially influence thermal transport and provide qualitative indications of its relative importance across different magnetic materials.
However, measurements of the total thermal conductivity alone cannot quantitatively disentangle how this coupling modifies phonon and magnon excitations and their scattering, or determine their respective contributions to heat transport.

To address these limitations, first principles based simulation approaches have been developed to resolve spin--phonon interactions at the microscopic level and connect them with thermal transport.
For antiferromagnetic NiO, microscopic calculations have resolved magnon--phonon scattering processes, quasiparticle lifetimes, and spin and lattice equilibration, revealing scattering channels involving two magnons with opposite polarizations and a phonon~\cite{maldonado2019microscopic}.
For ferromagnetic monolayer CrI$_3$, first principles calculations have quantified exchange mediated magnon--phonon scattering and shown that processes conserving magnon number are much stronger than those that change magnon number~\cite{cong2022exchange}.
At the transport level, first principles calculations combined with coupled magnon and phonon Boltzmann transport theory have enabled separate evaluation of the phonon and magnon contributions to nonelectronic thermal conductivity in ferromagnetic bcc Fe~\cite{pan2023ab}.
These approaches provide detailed microscopic information on scattering channels and their dependence on mode and wave vector.
However, the extensive electronic structure calculations required make direct simulations of coupled spin and lattice dynamics over large systems and long time scales computationally demanding.
The challenge becomes even greater at finite temperatures, where thermal lattice fluctuations and magnetic disorder require extensive sampling over many configurations~\cite{heine2019effect}.

As a complementary atomistic approach, spin-lattice dynamics (SLD) describes spin--phonon coupling through the dependence of magnetic interactions on the instantaneous atomic configuration.
Lattice vibrations modulate the distance dependent exchange interactions, thereby coupling the atomic and spin degrees of freedom~\cite{ma2008large,patra2023indirect}.
This real space formulation enables the coupled lattice and spin dynamics to evolve simultaneously at finite temperatures without explicitly evaluating mode specific magnon--phonon coupling matrix elements.
For ferromagnetic bcc Fe, SLD combined with spectral energy density (SED) analysis has been used to resolve phonon and magnon dispersions and scattering rates and to evaluate their respective contributions to thermal conductivity~\cite{wu2018magnon}.
Green--Kubo equilibrium atomistic and spin dynamics (GKEASD) further enables direct evaluation of phonon and magnon thermal transport while accounting for their mutual interactions~\cite{zhou2020atomistic}.
Nevertheless, existing SLD studies of thermal transport have focused predominantly on ferromagnetic systems, particularly bcc Fe, while direct comparisons of the effects of magnon--phonon coupling on thermal transport between ferromagnetic and antiferromagnetic materials within a unified atomistic framework remain scarce.
Extending SLD simulations to antiferromagnetic systems also introduces additional challenges, including the potentially nontrivial dependence of exchange interactions on interatomic distance beyond simple Bethe--Slater type descriptions~\cite{slater1930atomic,slater1930cohesion,sommerfeld1933handbuch}, as well as the mismatch between magnetic and chemical periodicities, which leads to distinct magnon spectra and available magnon--phonon scattering channels~\cite{streib2019magnon}.
It therefore remains unclear how differences in magnetic order translate into distinct spin--phonon coupling mechanisms and thermal transport behavior.

In this work, we use a unified SLD framework to investigate spin--phonon coupling and thermal transport in three magnetic monoxides with identical crystal structures but distinct magnetic order: antiferromagnetic NiO and MnO and ferromagnetic EuO.
To capture the coupling between spin and lattice degrees of freedom, the magnetic exchange interactions are modeled as power law functions of the interatomic distance.
Lattice thermal conductivity is evaluated using the Green--Kubo formalism, while phonon and magnon SED analyses are used to resolve the mode dependent spectral features and scattering behavior.
The comparison between MnO and NiO suggests that the spin induced change in thermal transport is sensitive to the relative phonon and magnon energy scales, with the stronger reduction in MnO coinciding with a greater spectral overlap between the two excitations.
For comparable phonon and magnon energy scales, the larger phonon--magnon scattering phase space in antiferromagnetic systems may contribute to the stronger reduction in lattice thermal conductivity observed in MnO than in EuO, although the quantitative scattering strength also depends on the coupling matrix elements.
These results provide microscopic insight into how phonon and magnon energy scales, magnetic order, and the available scattering pathways influence coupled spin-lattice thermal transport in magnetic materials.

\section{Methodology}
\subsection{Spin Lattice Dynamics Simulations}

We perform SLD simulations, in which the spin degrees of freedom are coupled to the lattice degrees of freedom, using the SPIN package implemented in LAMMPS~\cite{tranchida2018massively}.
Within this framework, magnetic effects are incorporated into classical molecular dynamics through a generalized Hamiltonian

\begin{equation}
    \mathcal{H}
    =
    \sum_{i=1}^{N}
    \frac{|\boldsymbol{p}_i|^2}{2m_i}
    +
    \sum_{i<j}^{N}
    V(r_{ij})
    +
    \mathcal{H}_{\mathrm{mag}} .
    \label{eq:SLD_Hamiltonian}
\end{equation}

The first term represents the kinetic energy of the atoms, while the second term describes the interionic interactions between the atoms.
The interionic interactions are modeled within the Born model of ionic solids, including long-range Coulomb interactions and short-range interactions represented by the Buckingham potential~\cite{lewis1985potential}.
The short-range Buckingham interaction is given by

\begin{equation}
    V_{\mathrm{Buck}}(r_{ij})
    =
    A_{ij}\exp\left(-B_{ij}r_{ij}\right)
    -
    \frac{C_{ij}}{r_{ij}^{6}},
\end{equation}
where $A_{ij}$, $B_{ij}$, and $C_{ij}$ are the Buckingham potential parameters for each ion pair.
The parameters used for NiO, MnO, and EuO are listed in Table~\ref{tab:buckingham}.
The final term, $\mathcal{H}_{\mathrm{mag}}$, represents the magnetic contribution to the Hamiltonian, which can in general include spin--spin exchange, magnetic anisotropy, Zeeman, dipolar, Dzyaloshinskii--Moriya, and magnetoelectric interactions.
In the present work, only isotropic spin--spin exchange interactions are included in the spin Hamiltonian.
For NiO and MnO, the easy-plane anisotropy associated with the $\{111\}$ planes is neglected because it is much weaker than the dominant isotropic exchange interactions~\cite{hutchings1972measurement,keffer1957problem}.
The even weaker in-plane easy-axis anisotropy along the $\langle 11\bar{2}\rangle$ directions is also omitted~\cite{hutchings1972measurement,keffer1957problem}.
For EuO, the weak cubic magnetocrystalline anisotropy, which favors the $\langle 111\rangle$ directions, is likewise neglected~\cite{miyata1967magnetocrystalline}.
The corresponding anisotropy energy scales and their comparison with the isotropic exchange interactions are discussed quantitatively in the main text.
The magnetic Hamiltonian is therefore written as
\begin{equation}
    \mathcal{H}_{\mathrm{mag}}
    =
    -\sum_{i<j}^{N}
    J(r_{ij})\,
    \boldsymbol{s}_i\cdot\boldsymbol{s}_j .
\end{equation}
This Heisenberg Hamiltonian describes the spin--spin exchange interaction, where $\boldsymbol{s}_i$ and $\boldsymbol{s}_j$ are the normalized spin vectors of atoms $i$ and $j$, respectively, and $J(r_{ij})$ is the exchange parameter, which depends on the interatomic distance $r_{ij}$.
With this sign convention, $J>0$ and $J<0$ correspond to ferromagnetic and antiferromagnetic couplings, respectively.

\begin{table}[!htb]
\centering
\caption{\textbf{Buckingham potential parameters used in the SLD simulations.}}
\label{tab:buckingham}
\begin{ruledtabular}
\begin{tabular}{ccccc}
Interaction &
$A$ (eV) &
$B$ (\AA$^{-1}$) &
$C$ (eV\,\AA$^{6}$) &
Ref. \\
\hline
Ni$^{2+}$--Ni$^{2+}$ & 0.0     & 1.0 & 0.0  & \multirow{2}{*}{\cite{lewis1985potential}} \\
Ni$^{2+}$--O$^{2-}$  & 754.9   & 3.1 & 0.0  & \\
Mn$^{2+}$--Mn$^{2+}$ & 0.0     & 1.0 & 0.0  & \multirow{2}{*}{\cite{lewis1985potential}} \\
Mn$^{2+}$--O$^{2-}$  & 832.7   & 3.1 & 0.0  & \\
Eu$^{2+}$--Eu$^{2+}$ & 1715.0  & 3.2 & 0.0  & \multirow{2}{*}{\cite{stritt1997study}} \\
Eu$^{2+}$--O$^{2-}$  & 5045.4  & 3.4 & 34.0 & \\
O$^{2-}$--O$^{2-}$   & 22764.3 & 6.7 & 27.9 & \\
\end{tabular}
\end{ruledtabular}
\end{table}

The central aspect of this simulation scheme is the introduction of a classical spin vector $\textbf{\textit{s}}_i$ for each atom $i$.
This enables the magnetic degrees of freedom to be treated explicitly alongside the atomic degrees of freedom, including the momentum $\textbf{\textit{p}}_i$ and position $\textbf{\textit{r}}_i$.
The equations of motion (EOM) can be derived from the Hamiltonian in Eq.~\ref{eq:SLD_Hamiltonian}:
\begin{equation}
    \frac{d\textbf{\textit{r}}_i}{dt}
    = \frac{\textbf{\textit{p}}_i}{m_i},
\end{equation}
\begin{equation}
    \frac{d\textbf{\textit{p}}_i}{dt}
    =
    \sum_{j\neq i}^{N}
    \left[
    \frac{dV(r_{ij})}{dr_{ij}}
    +
    \frac{dJ(r_{ij})}{dr_{ij}}
    \textbf{\textit{s}}_i\cdot\textbf{\textit{s}}_j
    \right]
    \textbf{\textit{e}}_{ij},
\end{equation}
\begin{equation}
    \frac{d\textbf{\textit{s}}_i}{dt}
    =
    \boldsymbol{\omega}_i
    \times
    \textbf{\textit{s}}_i.
\end{equation}

Both MD and SLD calculations were performed using a supercell size of $10 \times 10 \times 10$ for NiO, MnO, and EuO. 
The convergence of the calculated thermal conductivity with respect to supercell size was verified as described in Appendix~\ref{app:conv}.
The spin orientations were initialized randomly and subsequently relaxed through spin minimization until the magnetic torques converged, yielding an energetically minimized spin configuration.
Each system was first equilibrated for 200~ps in the isothermal--isobaric (NPT) ensemble with a time step of 1~fs at its respective target temperature and zero pressure.
During this stage, a Langevin thermostat and a Berendsen barostat were employed to control the temperature and pressure, respectively.
The barostat was then removed, and the system was further equilibrated for another 200~ps in the canonical (NVT) ensemble at the same target temperature.
These equilibration procedures were performed to prepare well-relaxed atomic and spin configurations for the subsequent production simulations.

\subsection{First-Principles Calculations of Magnetic Exchange Parameters}
\label{sec:exchange}
The exchange parameters $J_{ij}$ for NiO, MnO, and EuO were calculated using density functional theory (DFT) as implemented in the Vienna \textit{Ab initio} Simulation Package (VASP)~\cite{kresse1996efficient}, employing the projector augmented-wave (PAW) method~\cite{blochl1994projector}.
A plane-wave energy cutoff of 500~eV was used, with electronic energy and ionic force convergence criteria of $1 \times 10^{-6}$~eV and $0.01$~eV/\AA, respectively.
The Perdew--Burke--Ernzerhof generalized gradient approximation (PBE-GGA)~\cite{perdew1996generalized} was employed for the exchange--correlation functional, together with a $\Gamma$-centered \textbf{k}-point mesh of $6 \times 6 \times 6$.

For rocksalt NiO, MnO, and EuO, the exchange parameters were obtained by mapping the DFT total energies of three collinear magnetic configurations onto a Heisenberg Hamiltonian.
As illustrated in Fig.~\ref{fig:fig1}(a), $J_1$ and $J_2$ denote the first- and second-nearest-neighbor exchange interactions between magnetic ions, respectively.
For all three materials, the same three magnetic configurations were considered: a ferromagnetic configuration, a type-I antiferromagnetic configuration with alternating spin planes along the [001] direction, and a type-II antiferromagnetic configuration with alternating spin planes along the [111] direction.
These configurations provide linearly independent spin-correlation sums for the first- and second-nearest-neighbor pair classes, allowing $E_0$, $J_1$, and $J_2$ to be determined consistently.

For each magnetic configuration, the spin-correlation sums were evaluated as
\begin{equation}
C_{\alpha}
=
\sum_{\langle ij\rangle_{\alpha}} s_i s_j ,
\end{equation}
where $\alpha=1,2$ denotes the first- and second-nearest-neighbor pair classes, respectively, and $s_i=\pm1$ denotes the collinear spin orientation of magnetic ion $i$.
The DFT total energies were then mapped as
\begin{equation}
E
=
E_0
-
J_1 C_1
-
J_2 C_2 .
\end{equation}
The resulting $J_1$ and $J_2$ values are shown in Fig.~\ref{fig:fig1}(b)--(d).

To determine the distance dependence of the magnetic exchange interactions, the same supercells were uniformly scaled by factors of 0.96, 0.98, 1.00, 1.02, and 1.04 relative to the equilibrium lattice dimensions.
For each scaling factor, the DFT total energies of the corresponding magnetic configurations were recalculated, and the exchange parameters were extracted using the same energy-mapping procedure.
The resulting exchange parameters were fitted by the power-law relation~\cite{bloch1966103,fischer2009exchange}.
\begin{equation}
    J_{ij}(r_{ij})
    =
    J_{ij,0}
    \left(
    \frac{r_{ij}}{r_{ij,0}}
    \right)^{-n_{ij}},
    \label{method:s1}
\end{equation}
where $J_{ij,0}$ is the exchange parameter at the equilibrium pair distance $r_{ij,0}$, and $n_{ij}$ characterizes its sensitivity to the pair distance.

\subsection{Thermal Conductivity Calculations}
\label{sec:thermal}

The phonon and spin thermal conductivities were calculated using the Green--Kubo formalism from the corresponding heat-flux autocorrelation functions~\cite{kubo1957statistical}.
Following the NPT and NVT equilibration stages described above, the thermostat was removed and the systems were propagated in the microcanonical (NVE) ensemble for a 5~ns production trajectory.
The convergence of the calculated thermal conductivities with respect to the production trajectory length was verified as described in Appendix~\ref{app:conv}.

The phonon thermal conductivity along the $\alpha$ direction was evaluated as
\begin{equation}
\kappa_{\mathrm{ph}}^{\alpha}
=
\frac{V}{k_{\mathrm{B}}T^{2}}
\int_{0}^{\infty}
\left\langle
Q_{\mathrm{ph}}^{\alpha}(t)
Q_{\mathrm{ph}}^{\alpha}(0)
\right\rangle
\,\mathrm{d}t,
\end{equation}
where $k_{\mathrm{B}}$ is the Boltzmann constant, $T$ is the temperature, $V$ is the system volume, and $Q_{\mathrm{ph}}^{\alpha}$ is the phonon heat flux along the $\alpha$ direction.
The spin heat flux was evaluated from the time evolution of the pairwise exchange energy as
\begin{equation}
Q_{\mathrm{spin}}^{\alpha}
=
\frac{1}{V}
\sum_{i<j}
r_{ij}^{\alpha}
\frac{\mathrm{d}E_{ij}^{\mathrm{spin}}}{\mathrm{d}t},
\end{equation}
where
\begin{equation}
\frac{\mathrm{d}E_{ij}^{\mathrm{spin}}}{\mathrm{d}t}
=
-
J(r_{ij})
\left(
\dot{\bm{s}}_i\cdot\bm{s}_j
+
\bm{s}_i\cdot\dot{\bm{s}}_j
\right)
-
\dot{J}(r_{ij})
\bm{s}_i\cdot\bm{s}_j .
\end{equation}
In evaluating the spin heat flux, the contribution associated with the time variation of the exchange coefficient, $\dot{J}(r_{ij})\bm{s}_i\cdot\bm{s}_j$, was neglected~\cite{zhou2020atomistic}, giving
\begin{equation}
\frac{\mathrm{d}E_{ij}^{\mathrm{spin}}}{\mathrm{d}t}
\approx
-
J(r_{ij})
\left(
\dot{\bm{s}}_i\cdot\bm{s}_j
+
\bm{s}_i\cdot\dot{\bm{s}}_j
\right).
\end{equation}
Accordingly, the total thermal conductivity can be decomposed as
\begin{equation}
\kappa_{\mathrm{tot}}^{\alpha}
=
\kappa_{\mathrm{ph}}^{\alpha}
+
\kappa_{\mathrm{mag}}^{\alpha}
.
\end{equation}
Because classical molecular dynamics follows equipartition statistics and therefore overestimates the phonon heat capacity at low temperatures, a quantum heat-capacity correction was applied to the calculated phonon thermal conductivities of MnO and EuO at the temperatures considered in the main text (70 and 40~K, respectively)~\cite{bedoya2014computation}.
No quantum correction was applied to the magnon thermal conductivity because the simulations of MnO and EuO were performed at temperatures that are substantial fractions of their respective magnetic ordering temperatures.
The quantum-corrected phonon thermal conductivity was calculated as
\begin{equation}
\kappa_{\mathrm{ph,qc}}^{\alpha}
=
\kappa_{\mathrm{ph,cl}}^{\alpha}
\frac{C_V^{\mathrm{q}}}{C_V^{\mathrm{cl}}},
\end{equation}
where $\kappa_{\mathrm{ph,cl}}^{\alpha}$ is the phonon thermal conductivity obtained from the classical MD or SLD simulation, and $C_V^{\mathrm{q}}$ and $C_V^{\mathrm{cl}}$ are the quantum and classical phonon heat capacities, respectively.
The quantum phonon heat capacity was evaluated from the harmonic phonon density of states obtained from first-principles calculations using Bose--Einstein statistics,
\begin{equation}
C_V^{\mathrm{q}}(T)
=
k_{\mathrm{B}}
\int
g(\omega)
\left(
\frac{\hbar\omega}{k_{\mathrm{B}}T}
\right)^{2}
\frac{
\exp\left(\hbar\omega/k_{\mathrm{B}}T\right)
}{
\left[
\exp\left(\hbar\omega/k_{\mathrm{B}}T\right)-1
\right]^{2}
}
\,d\omega,
\end{equation}
where $g(\omega)$ is the phonon density of states.
The corresponding classical phonon heat capacity, $C_V^{\mathrm{cl}}$, was evaluated from the classical high-temperature limit using the same phonon density of states.
The same correction factor, $C_V^{\mathrm{q}}/C_V^{\mathrm{cl}}$, was applied to the phonon thermal conductivities obtained from both MD and SLD simulations at a given temperature to enable a consistent comparison of the spin-induced changes in thermal transport.

\subsection{Spectral Energy Density (SED) Calculations for Phonons and Magnons}
The phonon SED is calculated by summing over the basis atoms in the unit cell without considering projection onto eigenvectors~\cite{thomas2010predicting}
\begin{equation}
    \Phi(\mathbf{q},\omega)
    =
    \frac{1}{4\pi\tau_0 N_T}
    \sum_{\alpha}
    \sum_{b=1}^{n}
    m_b
    \left|
    \int_0^{\tau_0}
    \sum_{l=1}^{N_T}
    \dot{u}_{\alpha}(l,b,t)
    \exp\left[i\mathbf{q}\cdot\mathbf{r}_0(l)-i\omega t\right]
    \,\mathrm{d}t
    \right|^2 .
\end{equation}
Here, $\tau_0$ is the total MD simulation time, $N_T$ is the number of unit cells, $\dot{u}_{\alpha}(l,b,t)$ is the velocity of the $b$th basis atom in the $l$th unit cell along the $\alpha$ direction, $\mathbf{r}_0(l)$ is the equilibrium position of the $l$th unit cell, and $m_b$ is the atomic mass of the $b$th basis atom.
Similarly, the magnon SED is obtained from the time derivative of the transverse spin fluctuations, $\dot{S}_{\alpha}^{\perp}(l,b,t)$, where $S_{\alpha}^{\perp}(l,b,t)$ is the spin component of the $b$th magnetic basis atom in the $l$th unit cell along the transverse direction $\alpha$ relative to the equilibrium spin direction.
The time derivative $\dot{S}_{\alpha}^{\perp}(l,b,t)=\mathrm{d}S_{\alpha}^{\perp}(l,b,t)/\mathrm{d}t$ therefore describes the instantaneous transverse spin dynamics associated with magnon excitations.
The magnon SED is then obtained by summing over the $n_{\mathrm{mag}}$ magnetic basis atoms in each unit cell as
\begin{equation}
\Psi(\mathbf{k},\omega)
=
\frac{1}{4\pi\tau_0 N_T}
\sum_{\alpha}
\sum_{b=1}^{n_{\mathrm{mag}}}
\left|
\int_0^{\tau_0}
\sum_{l=1}^{N_T}
\dot{S}_{\alpha}^{\perp}(l,b,t)
\exp\left[
i\mathbf{k}\cdot\mathbf{r}_0(l)-i\omega t
\right]
\,\mathrm{d}t
\right|^2 .
\end{equation}
The phonon and magnon SED intensities, $\Phi(\mathbf{q},\omega)$ and $\Psi(\mathbf{k},\omega)$, are fitted using Lorentzian functions:
\begin{equation}
    \Phi(\mathbf{q},\omega)
    =
    \sum_{\nu}
    \frac{I_{\nu}}
    {
    \left[
    \frac{\omega-\omega_{\nu,c}(\mathbf{q})}
    {\gamma_{\nu}(\mathbf{q})}
    \right]^2
    +1
    },
\end{equation}
\begin{equation}
    \Psi(\mathbf{k},\omega)
    =
    \sum_{\mu}
    \frac{I_{\mu}}
    {
    \left[
    \frac{\omega-\omega_{\mu,c}(\mathbf{k})}
    {\gamma_{\mu}(\mathbf{k})}
    \right]^2
    +1
    }.
\end{equation}
Here, $\nu$ and $\mu$ denote the phonon and magnon branch indices, respectively. $I_{\nu}$ and $I_{\mu}$ are the peak amplitudes, $\omega_{\nu,c}$ and $\omega_{\mu,c}$ are the peak-center frequencies, and $\gamma_{\nu}$ and $\gamma_{\mu}$ are the corresponding half-widths at half-maximum. The phonon and magnon lifetimes are then calculated as $\tau=1/(2\gamma)$.

The NVE production trajectory used for the SED calculations was 800 ps with a time step of 1 fs, following 200 ps NPT and 200 ps NVT equilibration stages, with the equilibration and stability of the trajectory verified as described in Appendix~\ref{app:conv}.

\section{Results and Discussion}
\begin{figure}[!htb]

\includegraphics[width=\textwidth]{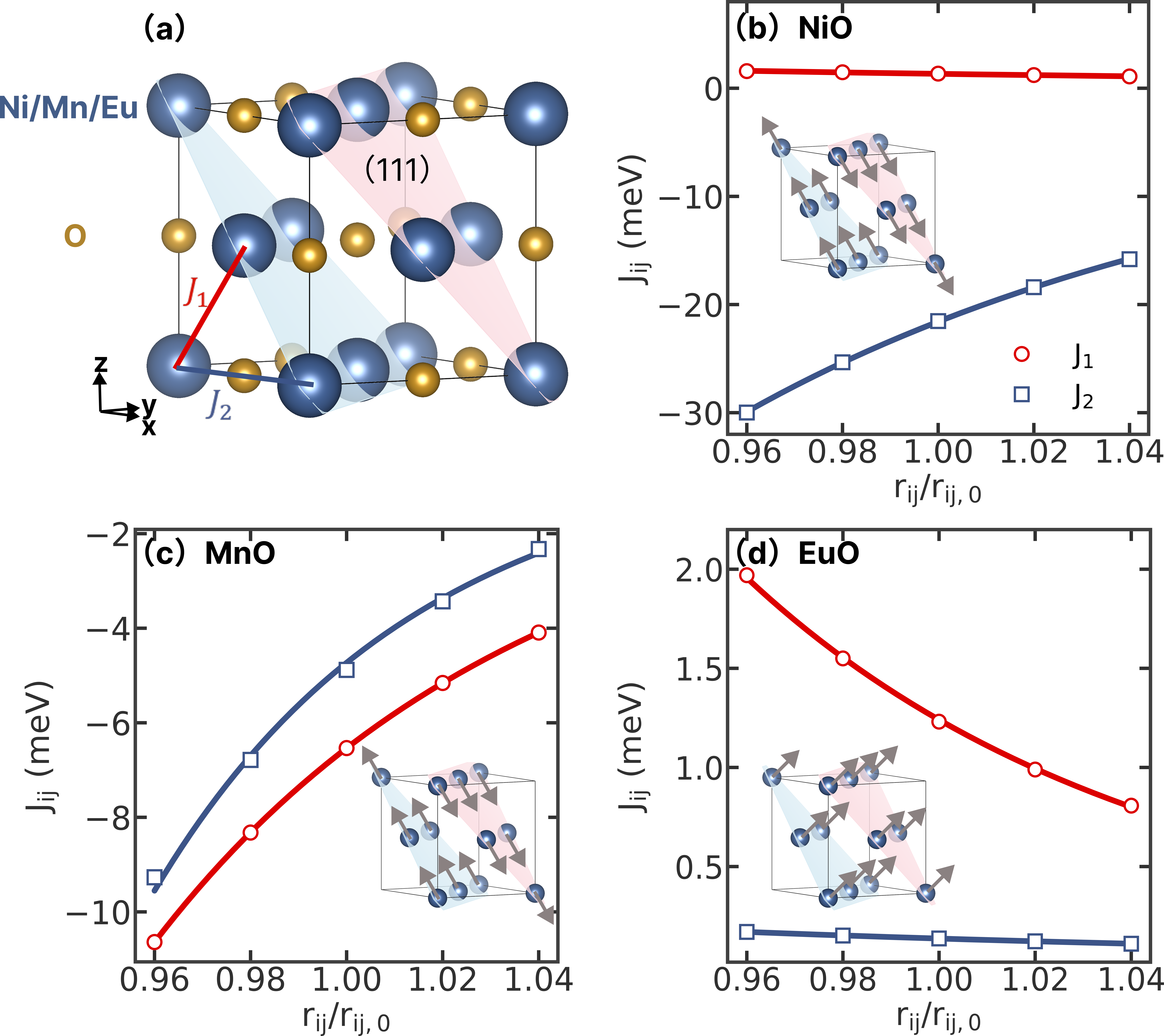}
\caption{\textbf{Exchange interactions in rocksalt NiO, MnO, and EuO.}
(a) Schematic of the rocksalt structure, showing the first- and second-nearest-neighbor exchange interactions, $J_1$ and $J_2$, between magnetic ions.
The shaded planes denote representative $\{111\}$ magnetic-ion planes associated with the AFII order in NiO and MnO.
(b)--(d) Exchange parameters $J_1$ and $J_2$ as functions of the normalized magnetic-ion pair distance $r_{ij}/r_{ij,0}$ for (b) NiO, (c) MnO, and (d) EuO, where $r_{ij,0}$ denotes the equilibrium pair distance.
Circles and squares denote the calculated $J_1$ and $J_2$, respectively, and the solid lines show the corresponding fits.
The insets illustrate the AFII magnetic order in NiO and MnO and the ferromagnetic order in EuO.}
\label{fig:fig1}
\end{figure}

MnO, NiO, and EuO share the rocksalt (halite) structure but exhibit different magnetic orders: MnO and NiO are antiferromagnetic type-II (AFII) systems dominated by oxygen-mediated superexchange~\cite{anderson1959new}, whereas EuO is ferromagnetic.
As shown in Fig.~\ref{fig:fig1}, the AFII structures of MnO and NiO consist of ferromagnetically aligned magnetic moments within the $\{111\}$ planes, with adjacent planes coupled antiferromagnetically along a $\langle 111\rangle$ direction~\cite{rezende2019introduction}.
The moments lie within the $\{111\}$ planes and, for the magnetic domains considered here, are oriented along a $\langle 11\bar{2}\rangle$ direction~\cite{rezende2019introduction}.
In contrast, EuO exhibits ferromagnetic order with parallel Eu moments, while its weak cubic magnetocrystalline anisotropy favors the $\langle 111\rangle$ directions~\cite{miyata1967magnetocrystalline}.
The exchange parameters $J_{ij}$ depend on both the interatomic distance $r_{ij}$ and local bonding geometry~\cite{kanamori1959superexchange,goodenough1955theory,anderson1950antiferromagnetism}.
For simplicity, the present SLD simulations neglect the explicit angular dependence of $J_{ij}$ and retain only its dependence on the magnetic-ion pair distance $r_{ij}$, as described as in Eq.~\ref{method:s1}.
The exchange parameters $J_{ij}$ obtained from first-principles calculations as a function of the normalized magnetic-ion pair distance $r_{ij}/r_{ij,0}$ for NiO, MnO, and EuO are shown in Fig.~\ref{fig:fig1}.
Details of the first-principles calculations and the extraction of the distance-dependent exchange parameters are described in Sec.~\ref{sec:exchange}.
The calculated equilibrium lattice constants of NiO, MnO, and EuO at 0~K are
4.20, 4.45, and 5.18~\AA, respectively.
These values are consistent with previous first-principles calculations~\cite{tran2006hybrid,an2011first}
and are in good agreement with the corresponding experimental values of
4.18, 4.45, and 5.15~\AA~\cite{gavriliuk2023first,kantor2005phase,guerci1966growth}.

The calculated equilibrium exchange parameters and fitted power-law exponents are compared with reported literature values in Table~\ref{tab:exchange_comparison}.
The corresponding equilibrium exchange parameters are $J_1=1.4$~meV and $J_2=-21.5$~meV for NiO, $J_1=-4.8$~meV and $J_2=-5.6$~meV for MnO, and $J_1=1.2$~meV and $J_2=0.1$~meV for EuO, where $J_1$ and $J_2$ denote the first- and second-nearest-neighbor exchange interactions, respectively.
In NiO, $J_1$ is relatively small, whereas the much larger negative $J_2$ dominates the AFII ordering.
In MnO, the negative $J_1$ and $J_2$ both favor antiferromagnetic ordering.
In EuO, the positive $J_1$ and weaker positive $J_2$ favor ferromagnetic ordering.
Overall, the equilibrium exchange parameters are in good agreement with previous results.
The fitted exponents are $n_1=5.5$ and $n_2=8.0$ for NiO, $n_1=12.0$ and $n_2=16.1$ for MnO, and $n_1=12.5$ and $n_2=5.4$ for EuO.
These values are of comparable magnitude to the reported distance sensitivities of the exchange interactions.
The exponents $n_1$ and $n_2$ quantify the relative distance sensitivity of the exchange interactions, whereas the absolute exchange-striction strength also depends on the magnitude of $J_{ij}$ through $|dJ_{ij}/dr_{ij}|=n_{ij}|J_{ij}|/r_{ij}$.

\begin{table*}[t]
\caption{
Comparison of the equilibrium exchange parameters and the corresponding fitted power-law exponents calculated in this work with reported literature values.
The MnO reference exponents are obtained by fitting the HSE results reported in Ref.~\cite{archer2011exchange} over the lattice parameter range of 4.0--4.5~\AA.
}
\label{tab:exchange_comparison}
\centering
\renewcommand{\arraystretch}{1}
\setlength{\tabcolsep}{5.5pt}
\begin{tabular}{lcccccccc}
\hline\hline
& \multicolumn{4}{c}{$J_1$}
& \multicolumn{4}{c}{$J_2$} \\
\cline{2-5}\cline{6-9}
Material
& $J_1$ (meV) & Ref. & $n_1$ & Ref.
& $J_2$ (meV) & Ref. & $n_2$ & Ref. \\
\hline

NiO
& 1.4
& 1.4~\cite{hutchings1972measurement}
& 5.5
& 3.7~\cite{zhang2006pressure}
& -21.5
& -19.0~\cite{hutchings1972measurement}
& 8.0
& 8.2~\cite{zhang2006pressure} \\

MnO
& -4.8
& -4.8~\cite{solovyev1998effective}
& 12.0
& 13.8~\cite{archer2011exchange}
& -5.6
& -5.6~\cite{solovyev1998effective}
& 16.1
& 11.3~\cite{archer2011exchange} \\

EuO
& 1.2
& 1.3~\cite{passell1976neutron}
& 12.5
& 17.0~\cite{mcwhan1966magnetic}
& 0.1
& 0.3~\cite{passell1976neutron}
& 5.4
& 10.0~\cite{sollinger2010exchange} \\

\hline\hline
\end{tabular}
\end{table*}

We first determine the magnetic transition temperatures of the three materials using SLD simulations. Detailed procedures for determining the magnetic transition temperatures are provided in Appendix~\ref{app:transition}.
The resulting N\'eel temperatures are $T_{\mathrm{N}}\approx424$~K for NiO and $T_{\mathrm{N}}\approx100$~K for MnO, while the Curie temperature of EuO is $T_{\mathrm{C}}\approx60$~K.
For comparison, atomistic spin dynamics (ASD) simulations~\cite{evans2014atomistic}, also described in Appendix~\ref{app:transition}, yield $T_{\mathrm{N}}\approx446$~K for NiO, $T_{\mathrm{N}}\approx112$~K for MnO, and $T_{\mathrm{C}}\approx66$~K for EuO.
The systematically higher transition temperatures obtained from ASD can be attributed primarily to the absence of thermal lattice expansion, which is included in SLD and reduces the exchange parameters as the lattice expands.
These values are in reasonable agreement with the experimentally reported transition temperatures of approximately 523~K for NiO~\cite{negovetic1973critical}, 118~K for MnO~\cite{kantor2005phase}, and 69~K for EuO~\cite{ahn2005preparation}.
The remaining discrepancy, particularly for NiO, may arise from the classical treatment of the spins and limitations of the simplified Heisenberg Hamiltonian, including the finite-range pairwise exchange parametrization and the neglect of additional magnetic interactions, such as magnetic anisotropy.
Comparable underestimation of the N\'eel temperature of NiO has also been reported in previous finite-temperature calculations based on Heisenberg models~\cite{fischer2009exchange}.

\begin{figure}[!htb]
\centering
\includegraphics[width=\textwidth]{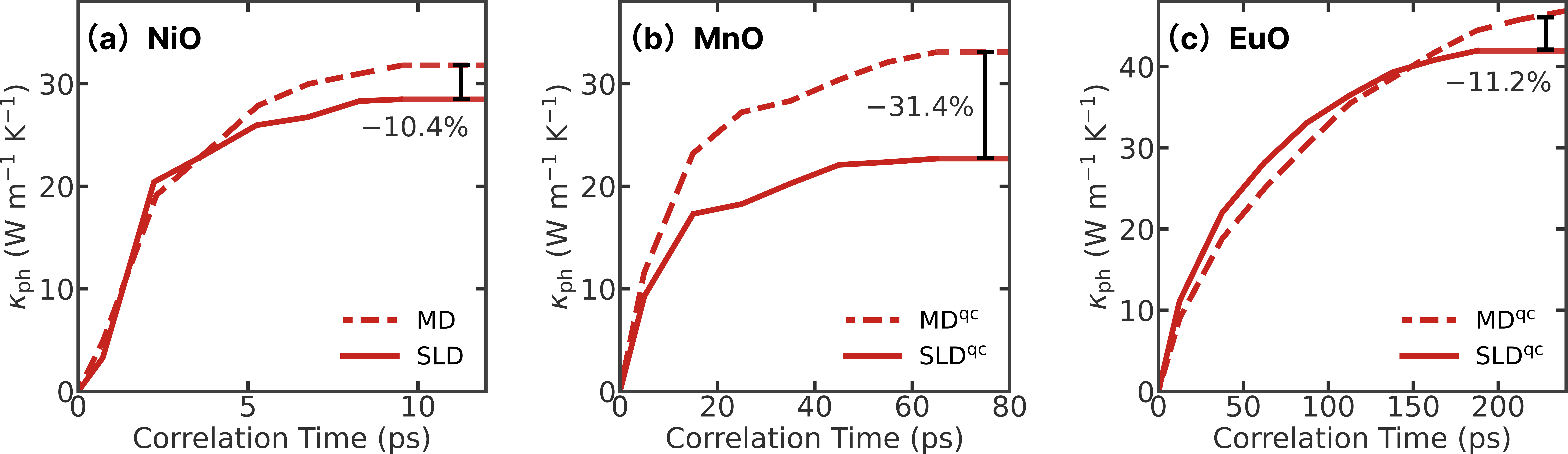}
\caption{\textbf{Phonon thermal conductivity of NiO, MnO, and EuO from molecular dynamics (MD) and spin--lattice dynamics (SLD) simulations.}
Phonon thermal conductivity as a function of correlation time for (a) NiO at 300~K, (b) MnO at 70~K, and (c) EuO at 40~K.
Including spin-lattice coupling reduces the converged phonon thermal conductivity by approximately 10.4\%, 31.4\%, and 11.2\% for NiO, MnO, and EuO, respectively.
Quantum heat-capacity corrections (qc) are applied to the MD and SLD results for MnO and EuO~\cite{bedoya2014computation}.}
\label{fig:fig2}
\end{figure}

We next examine the effect of spin--lattice coupling on thermal transport by comparing the lattice thermal conductivities obtained from SLD and conventional MD simulations.
Here we focus on the lattice thermal conductivity because the spin thermal conductivity is much smaller, as shown in Appendix~\ref{app:spinkappa}.
SLD explicitly includes the coupling between spin and lattice degrees of freedom, which is absent in conventional MD. 
Therefore, the magnitude of the relative change in lattice thermal conductivity upon including spin--lattice coupling can serve as a macroscopic indicator of the effective strength of spin--phonon coupling.
For a consistent comparison in the magnetically ordered phases, the transport calculations are performed at 300~K for NiO, 70~K for MnO, and 40~K for EuO, all near 70\% of their calculated transition temperatures.
A quantum heat-capacity correction, as described in Sec.~\ref{sec:thermal}, is applied to the calculated phonon thermal conductivities of MnO and EuO to account for the overestimation of the phonon heat capacity in classical MD at low temperatures~\cite{bedoya2014computation}.
Figure~\ref{fig:fig2} compares the thermal conductivities obtained from MD and SLD simulations using the Green--Kubo formalism~\cite{kubo1957statistical}.
Overall, the SLD results converge more rapidly with correlation time and yield lower thermal conductivities than the corresponding MD results.
This behavior is consistent with the inclusion of magnon–phonon coupling in SLD, which introduces an additional scattering channel, leading to faster convergence with correlation time and a reduced thermal conductivity.
For NiO at 300~K, Fig.~\ref{fig:fig2}(a) shows that the lattice thermal conductivity decreases from 31.8~$\mathrm{W\,m^{-1}\,K^{-1}}$ in MD to 28.5~$\mathrm{W\,m^{-1}\,K^{-1}}$ in SLD.
Including the spin thermal conductivity of 2.7~$\mathrm{W\,m^{-1}\,K^{-1}}$ at the same temperature obtained from the SLD simulation, as discussed in Appendix~\ref{app:spinkappa}, gives an approximate total thermal conductivity of 31.2~$\mathrm{W\,m^{-1}\,K^{-1}}$, which lies within the reported experimental range of approximately 20--37~$\mathrm{W\,m^{-1}\,K^{-1}}$ at room temperature~\cite{linnera2019lattice}.
For MnO at 70~K, Fig.~\ref{fig:fig2}(b) shows a larger decrease in the lattice thermal conductivity from 33.1 to 22.7~$\mathrm{W\,m^{-1}\,K^{-1}}$.
Including the spin contribution of 1.4~$\mathrm{W\,m^{-1}\,K^{-1}}$ gives an approximate total thermal conductivity of 24.1~$\mathrm{W\,m^{-1}\,K^{-1}}$, in close agreement with the experimental value of 24.7~$\mathrm{W\,m^{-1}\,K^{-1}}$~\cite{slack1958thermal}.
For EuO at 40~K, Fig.~\ref{fig:fig2}(c) shows that the lattice thermal conductivity decreases from 47.3 to 42.0~$\mathrm{W\,m^{-1}\,K^{-1}}$.
Including the spin contribution of 2.0~$\mathrm{W\,m^{-1}\,K^{-1}}$ gives an approximate total thermal conductivity of 44.0~$\mathrm{W\,m^{-1}\,K^{-1}}$, compared with the experimental value of 52.3~$\mathrm{W\,m^{-1}\,K^{-1}}$~\cite{martin1972thermal}.
The remaining discrepancies in the absolute thermal conductivities at low temperatures may partly arise from the simplified quantum correction employed here, which accounts for the quantum reduction of the phonon heat capacity while retaining the phonon relaxation times sampled from classical dynamics.
Classical statistics can overpopulate high-frequency phonons at low temperatures and thereby modify anharmonic scattering and phonon lifetimes, so correcting the heat capacity alone does not necessarily recover the fully quantum thermal conductivity~\cite{turney2009assessing}.
Relative to conventional MD, the inclusion of spin--lattice coupling in SLD reduces the lattice thermal conductivity by approximately 10.4\%, 31.4\%, and 11.2\% for NiO, MnO, and EuO, respectively.
The substantially larger reduction in MnO therefore indicates stronger magnon--phonon scattering.

\begin{figure}[!htb]
\centering
\includegraphics[width=\textwidth]{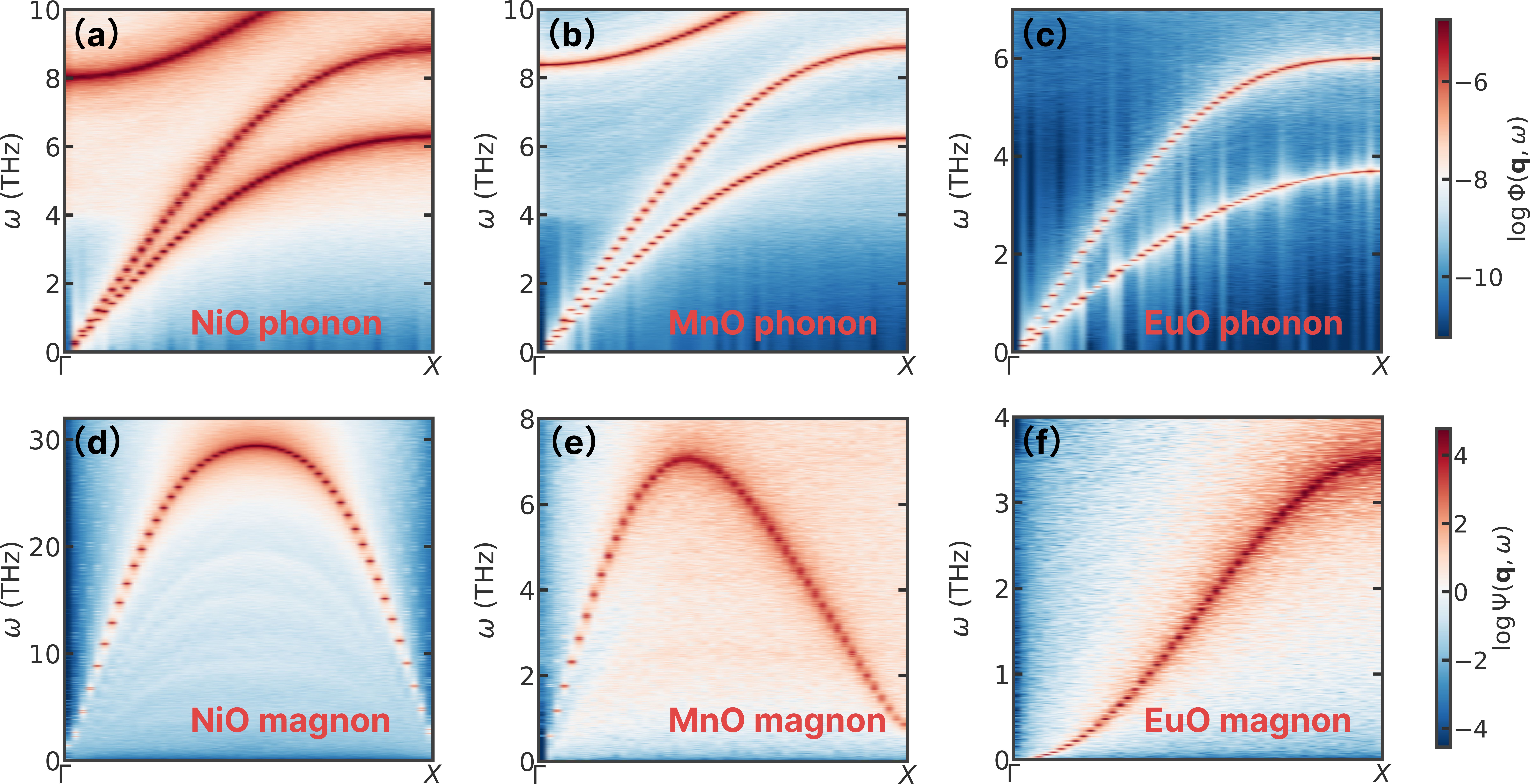}
\caption{\textbf{Phonon and magnon spectral energy densities (SED) of NiO, MnO, and EuO.}
(a)--(c) Phonon spectral energy density $\log \Phi(\mathbf{q},\omega)$ of NiO, MnO, and EuO, respectively, along the $\Gamma$--$X$ direction defined based on the chemical lattice, evaluated at the corresponding thermal-transport temperatures.
(d)--(f) Magnon spectral energy density $\log \Psi(\mathbf{q},\omega)$ along the same reciprocal-space, evaluated at 10~K.
The spectra reveal distinct phonon--magnon energy scales and contrasting magnon dispersion characteristics among the three materials.}
\label{fig:fig3}
\end{figure}

To probe the microscopic origin of this enhanced phonon-magnon scattering in MnO, we next examine the phonon and magnon excitation spectra using SED analysis~\cite{thomas2010predicting}.
The phonon SEDs are evaluated at the corresponding transport temperatures, whereas the magnon SEDs are calculated at 10~K to resolve the magnon dispersions more clearly, as thermal spin fluctuations near 70\% of the magnetic transition temperatures substantially broaden the magnon spectral features.
The three materials exhibit markedly different relative energy scales of the lattice and spin excitations.
For NiO, the two transverse acoustic (TA) phonon branches reach approximately 6.0~THz and the longitudinal acoustic (LA) phonon branch approximately 9.0~THz along the $\Gamma$--$X$ direction in Fig.~\ref{fig:fig3}(a), whereas the magnon branch reaches a maximum of approximately 30.0~THz in Fig.~\ref{fig:fig3}(d).
The nonmonotonic magnon dispersion reflects zone folding associated with the different magnetic and chemical translational periodicities, with the reciprocal-space path defined relative to the chemical lattice.
By comparison, MnO and EuO exhibit a much closer match between the magnon and acoustic-phonon energy scales.
In MnO, the TA and LA phonon branches extend to approximately 6.0 and 9.0~THz, respectively, while the magnon spectrum reaches approximately 7.0~THz, as shown in Fig.~\ref{fig:fig3}(b) and (e).
The degeneracy of the two magnon modes in NiO and MnO arises from the neglect of magnetic anisotropy in the present isotropic Heisenberg model.
For NiO, the reported easy-plane and in-plane easy-axis anisotropies correspond to energy scales of approximately $0.09$ and $0.005$~meV per Ni, respectively~\cite{hutchings1972measurement}, which are much smaller than the dominant exchange interaction $|J_2|\approx21.5$~meV.
For MnO, the corresponding easy-plane and in-plane anisotropy energy scales are approximately $0.14$ and $4\times10^{-4}$~meV per Mn, respectively~\cite{keffer1957problem}, compared with $|J_2|\approx5.6$~meV.
In EuO, the corresponding frequencies are approximately 4.0 and 6.0~THz for the TA and LA phonon branches and 3.5~THz for the magnon spectrum, as shown in Fig.~\ref{fig:fig3}(c) and (f).
Unlike the folded magnon dispersions in antiferromagnetic NiO and MnO, the EuO magnon branch increases monotonically along $\Gamma$--$X$, reflecting the identical magnetic and chemical translational periodicities in the ferromagnetic state.
Moreover, near $\Gamma$, both antiferromagnetic magnons and acoustic phonons exhibit linear dispersions, whereas the ferromagnetic magnon branch is quadratic, resulting in closer spectral matching in the antiferromagnetic systems.
Together, these differences in magnon dispersion and phonon--magnon energy overlap may influence the strength of spin-induced phonon scattering.
We therefore examine this effect more directly through the SED linewidths and lifetimes of the TA and LA phonon branches.

\begin{figure}[!htb]
\centering
\includegraphics[width=\textwidth]{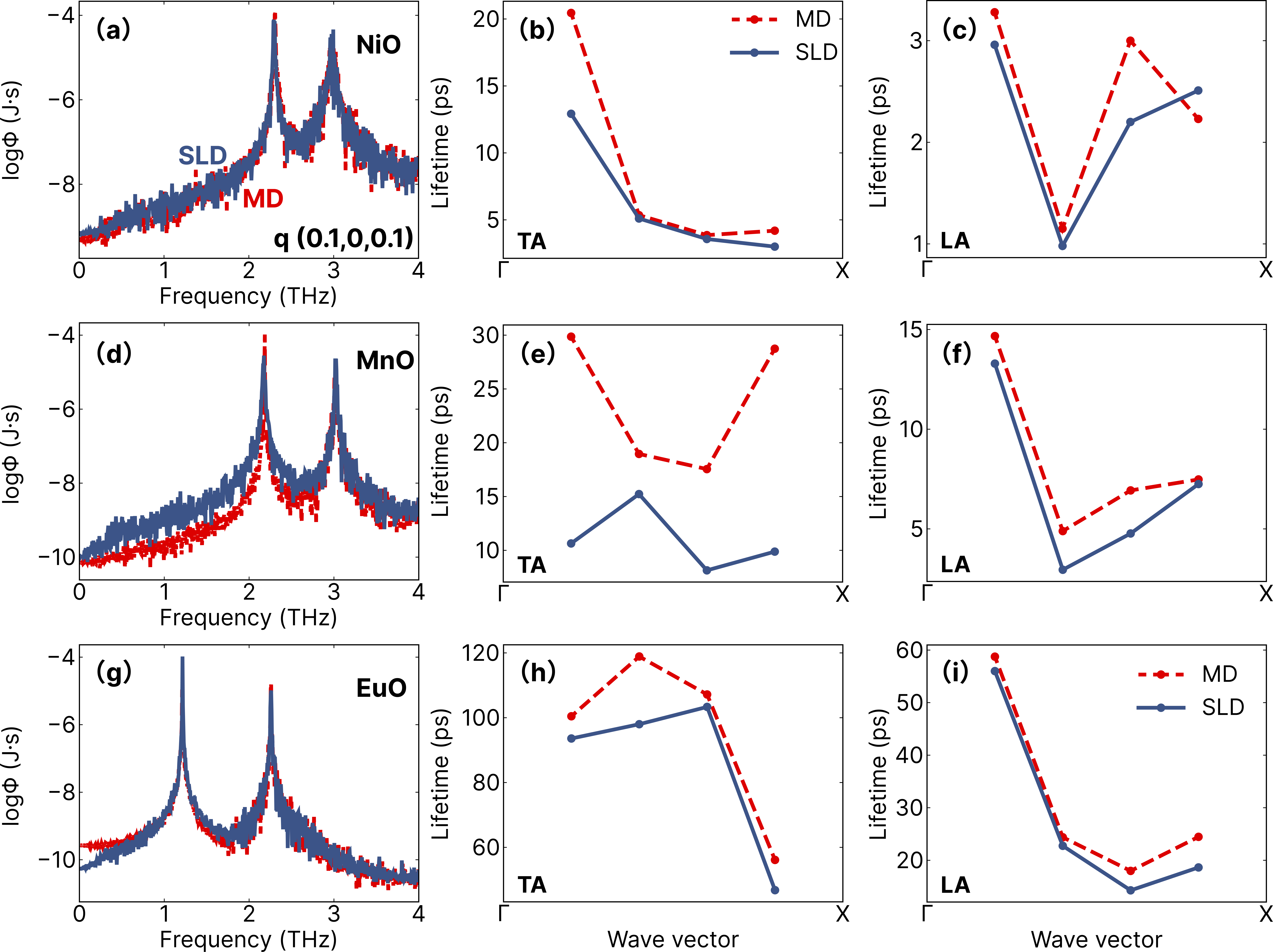}
\caption{\textbf{Effect of spin--lattice coupling on phonon spectra and lifetimes in NiO, MnO, and EuO.}
(a), (d), and (g) Phonon spectral energy density $\log\Phi(\omega)$ obtained from MD and SLD at $\mathbf{q}=(0.1,0,0.1)$ for NiO, MnO, and EuO, respectively.
(b), (e), and (h) Lifetimes of the transverse acoustic (TA) phonons along the $\Gamma$--$X$ direction obtained from MD and SLD.
(c), (f), and (i) Corresponding lifetimes of the longitudinal acoustic (LA) phonons.
The reduction of phonon lifetimes in SLD relative to MD reflects the additional phonon scattering induced by spin--lattice coupling, with the most pronounced effect observed in MnO.}
\label{fig:fig4}
\end{figure}

Figure~\ref{fig:fig4}(a), (d), and (g) compare representative phonon SEDs from MD and SLD simulations at $\mathbf{q}=(0.1,0,0.1)$ for NiO, MnO, and EuO, respectively, at the corresponding transport temperatures.
This wave vector was selected as a representative point near $\Gamma$, where the linear acoustic-phonon and antiferromagnetic-magnon dispersions provide favorable spectral overlap for phonon--magnon scattering.
Including spin dynamics produces pronounced broadening of the TA peak in MnO, while the spectra of NiO and EuO show little change.
To quantify this broadening more systematically, Lorentzian fits to the TA and LA peaks were used to extract the phonon lifetimes along the $\Gamma$--$X$ direction.
The extracted TA lifetimes along the $\Gamma$--$X$ direction are shown in Fig.~\ref{fig:fig4}(b), (e), and (h), and the LA lifetimes in Fig.~\ref{fig:fig4}(c), (f), and (i).
MnO exhibits a pronounced reduction in the TA lifetime and a weaker reduction in the LA lifetime, whereas both branches in NiO and EuO are only weakly affected by the inclusion of spin dynamics.
Together with the excitation spectra in Fig.~\ref{fig:fig3}, these trends point to an important role of the relative phonon--magnon energy scales.
Within MnO, the TA branch is closer in energy to the magnon spectrum and exhibits a larger lifetime reduction than the LA branch.
Similarly, compared with antiferromagnetic NiO, MnO exhibits closer phonon--magnon energy scales together with a much stronger reduction in phonon lifetime.
These comparisons suggest that closer energetic matching favors stronger spin-induced phonon scattering.
However, EuO also exhibits comparable phonon and magnon energy scales but only weak lifetime changes, indicating that energetic matching alone is insufficient to explain the enhanced scattering in MnO.
This contrast suggests that magnetic order itself may play an additional role in governing the scattering processes, which we examine below.

\begin{figure}[!htb]
\centering
\includegraphics[width=0.8\textwidth]{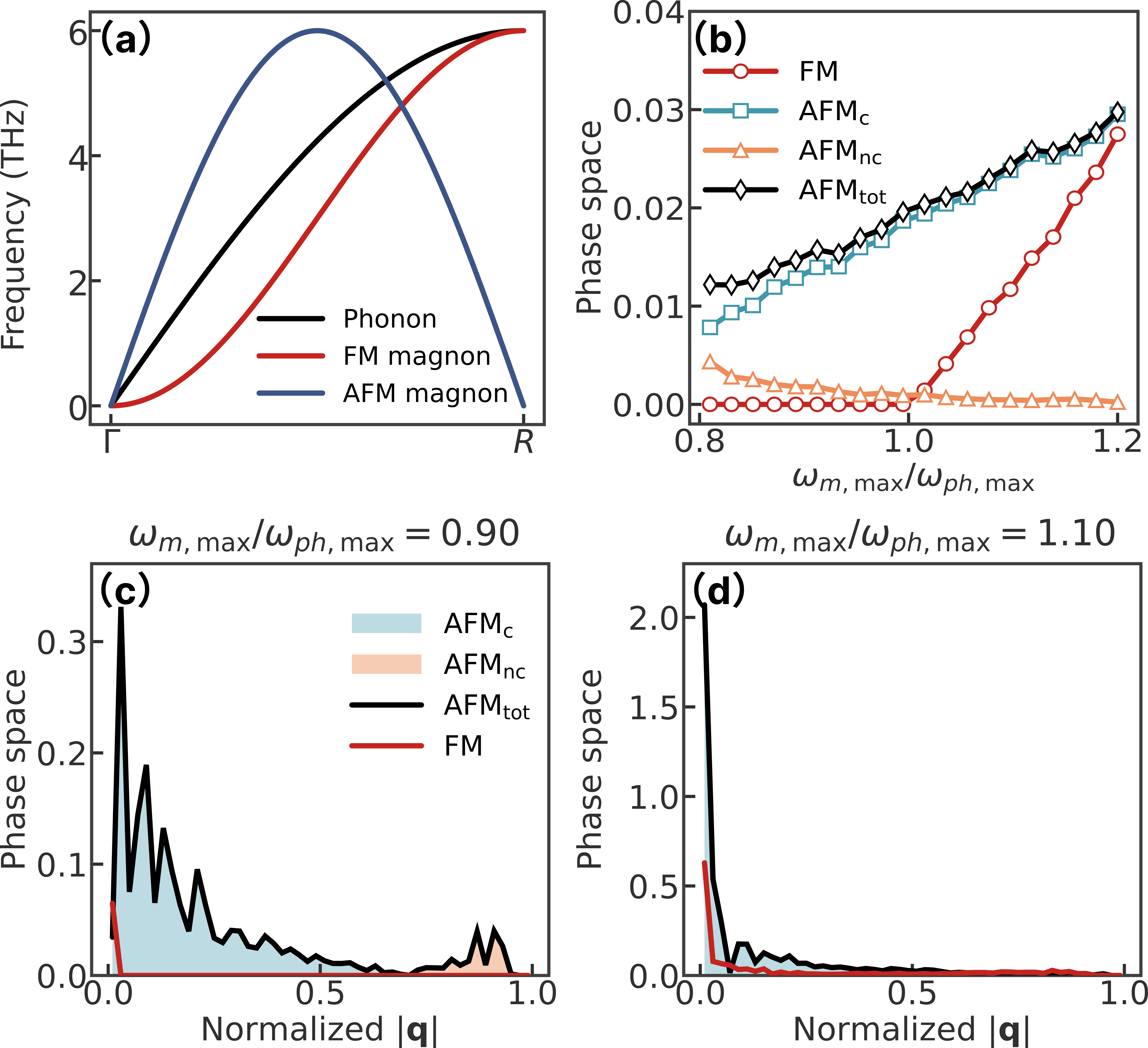}
\caption{
\textbf{Comparison of the magnon--phonon scattering phase space for ferromagnetic (FM) and antiferromagnetic (AFM) systems.}
(a) Model phonon, FM magnon, and AFM magnon dispersions plotted along the $\Gamma \rightarrow R$ direction for $\omega_{m,\max}/\omega_{ph,\max}=1.00$.
(b) Momentum-integrated kinematic phase space as a function of the ratio between the maximum magnon and phonon frequencies, $\omega_{m,\max}/\omega_{ph,\max}$.
For the AFM case, the magnon-number-conserving ($\mathrm{AFM}_{\mathrm{c}}$) and magnon-number-nonconserving ($\mathrm{AFM}_{\mathrm{nc}}$) contributions are shown separately, together with their sum ($\mathrm{AFM}_{\mathrm{tot}}$).
(c),(d) Momentum-resolved phase-space distributions as a function of normalized $|\mathbf{q}|$ for $\omega_{m,\max}/\omega_{ph,\max}=0.90$ and $1.10$, respectively.
The shaded regions indicate the conserving and nonconserving AFM contributions, while the solid black and red curves denote the total AFM and FM phase spaces, respectively.
}
\label{fig:fig5}
\end{figure}

To isolate the effect of magnetic order on the phonon--magnon scattering phase space, we construct a minimal model for ferromagnetic (FM) and antiferromagnetic (AFM) systems.
The model dispersions are chosen to capture the generic long-wavelength behavior of acoustic phonons and FM and AFM magnons while preserving the lattice periodicity and finite bandwidth across the Brillouin zone, as shown in Fig.~\ref{fig:fig5}(a).
As a modeling assumption, the acoustic phonons are represented by a single effective branch on a three-dimensional simple-cubic lattice with nearest-neighbor harmonic interactions.
Within this model, standard harmonic lattice dynamics~\cite{born1996dynamical} gives
\begin{equation}
\omega_{\mathrm{ph}}(\mathbf{q})
=
\omega_{\mathrm{ph,max}}
\left[
\frac{
\sin^2(q_x/2)+
\sin^2(q_y/2)+
\sin^2(q_z/2)
}{3}
\right]^{1/2},
\end{equation}
where the lattice constant is set to unity.
In the long-wavelength limit near $\Gamma$, this dispersion reduces to the expected linear behavior 
$\omega_{\mathrm{ph}}(\mathbf{q})\propto|\mathbf{q}|$~\cite{born1996dynamical}.
For the FM system, we assume an isotropic nearest-neighbor Heisenberg ferromagnet on the same simple-cubic lattice.
Within linear spin-wave theory, the Holstein--Primakoff treatment~\cite{holstein1940field} gives
\begin{equation}
\omega_{\mathrm{m}}^{\mathrm{FM}}(\mathbf{k})
=
\omega_{\mathrm{m,max}}
\frac{
3-\cos k_x-\cos k_y-\cos k_z
}{6},
\label{eq:FM}
\end{equation}
where the normalization is chosen such that the maximum magnon frequency is $\omega_{\mathrm{m,max}}$.
Near $\Gamma$, this dispersion exhibits the characteristic quadratic behavior
$\omega_{\mathrm{m}}^{\mathrm{FM}}(\mathbf{k})\propto|\mathbf{k}|^2$~\cite{holstein1940field}.
For the AFM system, we assume a bipartite nearest-neighbor Heisenberg antiferromagnet on the same simple-cubic lattice.
Within two-sublattice linear spin-wave theory~\cite{kubo1952spin}, the magnon dispersion is
\begin{equation}
\omega_{\mathrm{m}}^{\mathrm{AFM}}(\mathbf{k})
=
\omega_{\mathrm{m,max}}
\sqrt{1-\gamma_{\mathbf{k}}^2},
\qquad
\gamma_{\mathbf{k}}
=
\frac{\cos k_x+\cos k_y+\cos k_z}{3},
\end{equation}
where the normalization is chosen such that the maximum magnon frequency is $\omega_{\mathrm{m,max}}$.
Near $\Gamma$, this dispersion exhibits the characteristic linear behavior
$\omega_{\mathrm{m}}^{\mathrm{AFM}}(\mathbf{k})\propto|\mathbf{k}|$~\cite{kubo1952spin}.

In the FM case, the lowest-order exchange-mediated phonon--magnon scattering process conserves the magnon number~\cite{streib2019magnon}.
This can be understood physically from the conservation of spin angular momentum.
The process can be written as
\begin{equation}
\mathrm{ph}(\mathbf{q};0)
+
\mathrm{m}(\mathbf{k};s_m)
\leftrightarrow
\mathrm{m}(\mathbf{k}';s_m),
\end{equation}
where the second argument denotes the spin angular momentum associated with each excitation.
Within the present model, the phonon carries no spin angular momentum, while all magnons for a given FM ground state carry the same spin angular momentum $s_m$ relative to the equilibrium magnetization, with $s_m=-\hbar$ for the convention adopted here.

In contrast, the two-sublattice spin-wave structure of the AFM state supports two magnon modes with opposite spin angular momenta, labeled by $\sigma=\pm$, with $s_{m,\sigma}=\sigma\hbar$.
For magnon-number-conserving scattering, spin-angular-momentum conservation requires the incoming and outgoing magnons to carry the same spin angular momentum.
The corresponding processes can therefore be written as
\begin{equation}
\mathrm{ph}(\mathbf{q};0)
+
\mathrm{m}_{\sigma}(\mathbf{k};s_{m,\sigma})
\leftrightarrow
\mathrm{m}_{\sigma}(\mathbf{k}';s_{m,\sigma}),
\qquad
\sigma=\pm.
\end{equation}
Unlike in the FM case, however, the presence of two oppositely polarized magnon modes also allows magnon-pair creation and annihilation~\cite{streib2019magnon},
\begin{equation}
\mathrm{ph}(\mathbf{q};0)
\leftrightarrow
\mathrm{m}_{+}(\mathbf{k};s_{m,+})
+
\mathrm{m}_{-}(\mathbf{k}';s_{m,-}),
\end{equation}
where $s_{m,+}=+\hbar$ and $s_{m,-}=-\hbar$.
Although this process changes the magnon number by two, the two magnons carry opposite spin angular momenta, such that $s_{m,+}+s_{m,-}=0$, and spin angular momentum remains conserved.
The microscopic origin of these FM and AFM phonon--magnon scattering processes is derived in terms of magnon creation and annihilation operators in Appendix~\ref{app:scattering_pathways}.

Figure~\ref{fig:fig5}(b) compares the total phonon--magnon scattering phase space in the FM and AFM models as a function of
$R=\omega_{\mathrm{m,max}}/\omega_{\mathrm{ph,max}}$.
For an allowed scattering channel $\lambda$, the total phase space is defined as
\begin{equation}
\mathcal{P}_{\lambda}(R)
=
\frac{1}{N_qN_k}
\sum_{\mathbf{q},\mathbf{k}}
\delta
\left[
\Delta\omega_{\lambda}(\mathbf{q},\mathbf{k};R)
\right],
\label{eq:phase_space_total}
\end{equation}
where $\lambda$ labels the scattering channel, $\mathbf{q}$ and $\mathbf{k}$ denote the phonon and magnon wave vectors, respectively, and $N_q$ and $N_k$ are the numbers of sampled phonon and magnon wave vectors.
The quantity $\Delta\omega_{\lambda}$ is the energy mismatch of the corresponding scattering process, and the Dirac delta function enforces energy conservation.
Crystal-momentum conservation is imposed explicitly through the relation between the initial- and final-state wave vectors modulo a reciprocal lattice vector $\mathbf{G}$.
Spin-angular-momentum conservation is imposed by including only the allowed scattering channels introduced above.
As derived in Appendix~\ref{app:phase_space}, the FM phase space exhibits a threshold at $R_{\mathrm{c}}=1$, vanishing for $R<1$ and increasing rapidly for $R>1$.
In contrast, the AFM phase space remains nonzero throughout the range considered and is larger than the FM phase space, particularly for $R<1$.
This difference in the available phonon--magnon scattering phase space helps explain the weaker spin-induced phonon scattering in EuO than in MnO.
In addition, the AFM phase space is dominated by the magnon-number-conserving channel, while the nonconserving contribution is smaller.
To examine which phonon wave vectors contribute to the total phase space, we define the phonon-wave-vector-resolved phase space as
\begin{equation}
\mathcal{P}_{\lambda}(\mathbf{q};R)
=
\frac{1}{N_k}
\sum_{\mathbf{k}}
\delta
\left[
\Delta\omega_{\lambda}(\mathbf{q},\mathbf{k};R)
\right].
\label{eq:phase_space_q}
\end{equation}
For a given phonon wave vector $\mathbf{q}$, $\mathcal{P}_{\lambda}(\mathbf{q};R)$ measures the density of magnon states available for scattering with that phonon while satisfying the corresponding conservation laws.
The total phase space in Eq.~\ref{eq:phase_space_total} is obtained by averaging $\mathcal{P}_{\lambda}(\mathbf{q};R)$ over the phonon Brillouin zone.
Figures~\ref{fig:fig5}(c) and \ref{fig:fig5}(d) show this phonon-wave-vector-resolved phonon--magnon scattering phase space, radially averaged and plotted as a function of normalized $|\mathbf{q}|$, at $R=0.9$ and $R=1.1$, respectively.
At $R=0.9$, the FM phase space is strongly suppressed, whereas a finite phase space remains in the AFM case.
The AFM magnon-number-conserving contribution is mainly distributed at smaller $|\mathbf{q}|$, while the magnon-number-nonconserving contribution extends over a broader momentum range.
At $R=1.1$, the FM phase space becomes finite and increases, while the AFM phase space remains larger over much of the momentum range.
The magnon-number-nonconserving contribution is further reduced at this larger value of $R$, leaving the AFM phase space primarily determined by the magnon-number-conserving channel.
Overall, when the phonon and magnon energy scales are comparable, the larger scattering phase space in the AFM model provides more kinematically allowed channels and may therefore favor stronger phonon--magnon scattering than in the FM case.

It should be emphasized that the available scattering phase space alone does not determine the phonon--magnon scattering rate, which also depends on the corresponding interaction matrix elements~\cite{streib2019magnon,ruckriegel2014magnetoelastic,liu2017magnon}.
In the present SLD model, the exchange-mediated spin--lattice coupling originates from the distance dependence of the exchange interaction, with its strength related to $\partial J_{ij}/\partial r_{ij}$.
The mode-resolved scattering matrix elements further depend on the phonon and magnon eigenmodes involved.
Therefore, similar scattering phase spaces can still lead to different scattering rates because of differences in coupling strength and mode character.
Since these mode-resolved matrix elements are not explicitly evaluated here, the phase-space analysis should be interpreted as a qualitative measure of the available scattering channels rather than a quantitative prediction of the scattering rates.

\section{Conclusion}
In summary, we applied first-principles-informed spin--lattice dynamics simulations to investigate phonon--magnon interactions and thermal transport in antiferromagnetic NiO and MnO and ferromagnetic EuO.
By combining Green--Kubo thermal-transport calculations with phonon and magnon spectral energy density analyses, we found that spin dynamics reduce the phonon thermal conductivity and lifetimes, with the strongest effect observed in MnO.
The comparison among the three materials shows that the magnitude of spin-induced phonon scattering depends sensitively on the relative phonon and magnon energy scales.
A simplified scattering-phase-space model further shows that, for comparable excitation energy scales, the AFM state provides a larger phonon--magnon scattering phase space than the FM state, offering a microscopic explanation for the enhanced scattering observed in MnO.
These results highlight the combined roles of excitation energy scales and magnetic order in governing phonon--magnon scattering and provide guidelines for engineering thermal transport in magnetic materials.

\begin{acknowledgments}
This work is based on research partially supported by the U.S. Army Research Office under award number W911NF2310188 and a seed grant provided by the Institute of Energy Efficiency at the University of California, Santa Barbara (UCSB). This work used Stampede3 at the Texas Advanced Computing Center (TACC) through allocation MAT200011 from the Advanced Cyberinfrastructure Coordination Ecosystem: Services \& Support (ACCESS) program, which is supported by National Science Foundation grants 2138259, 2138286, 2138307, 2137603, and 2138296. Use was also made of computational facilities purchased with funds from the National Science Foundation (award number CNS-1725797) and administered by the Center for Scientific Computing (CSC) at the University of California, Santa Barbara (UCSB). The CSC is supported by the California NanoSystems Institute and the Materials Research Science and Engineering Center (MRSEC; NSF DMR-2308708) at UCSB. 

\end{acknowledgments}

\appendix

\section{Convergence Tests}
\label{app:conv}

\begin{figure}[!htb]
\centering
\includegraphics[width=0.5\textwidth]{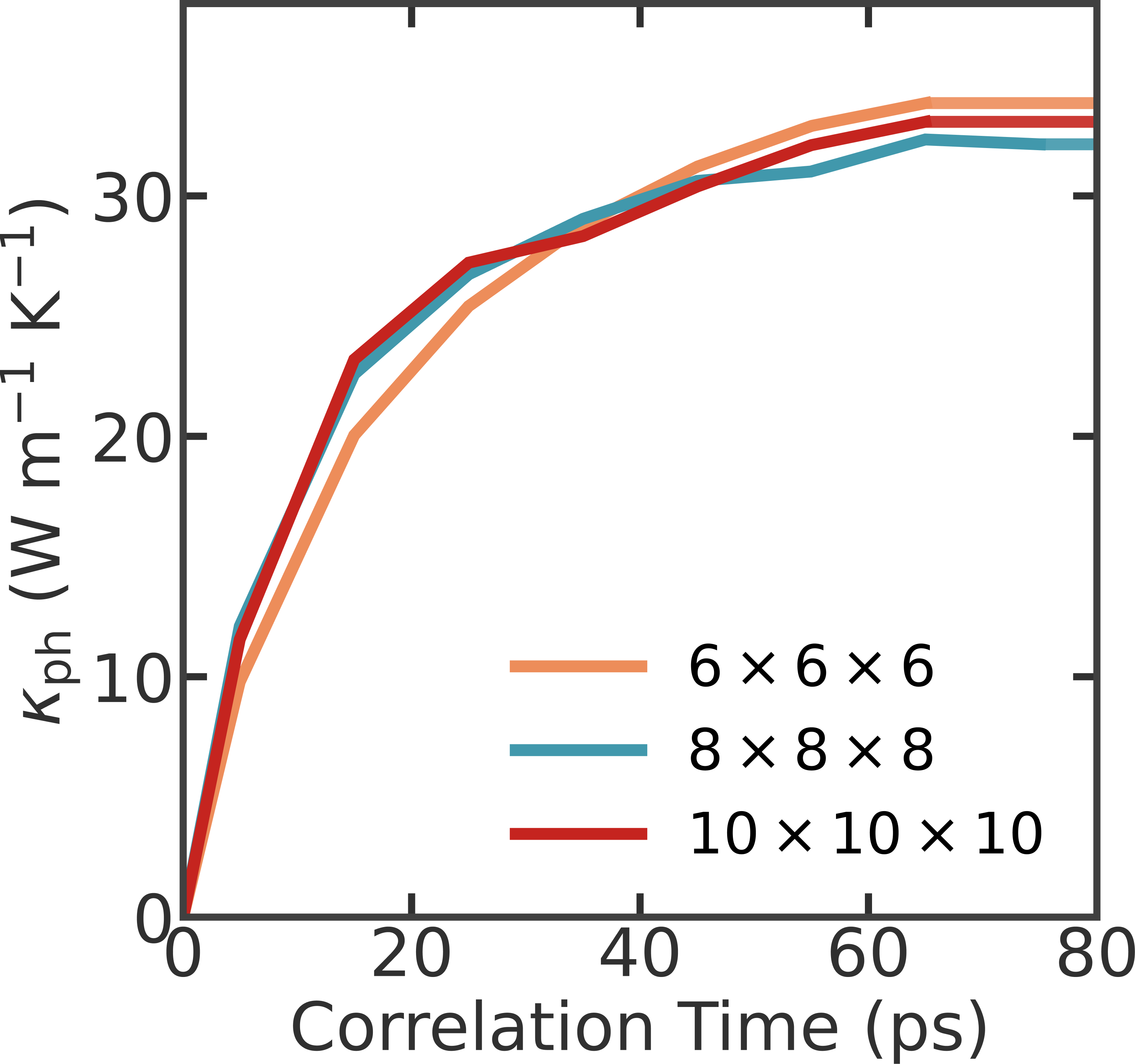}
\caption{Supercell-size convergence of the quantum-corrected lattice thermal conductivity, $\kappa_{\mathrm{ph}}$, of MnO at 70 K obtained from MD simulations. Results are shown for $6\times6\times6$, $8\times8\times8$, and $10\times10\times10$ supercells as a function of correlation time. The three calculations converge to comparable thermal conductivity values, indicating weak supercell-size dependence over the investigated size range.}
\label{fig:supercell}
\end{figure}

To examine the convergence with respect to supercell size, we calculated the lattice thermal conductivity of MnO at 70 K using MD simulations with $6\times6\times6$, $8\times8\times8$, and $10\times10\times10$ supercells.
As shown in Fig.~\ref{fig:supercell}, the thermal conductivity curves obtained using the three supercell sizes exhibit similar correlation-time dependence and converge to comparable values, indicating that the calculated lattice thermal conductivity is sufficiently converged with respect to the simulation cell size.

\begin{figure}[!htb]
\centering
\includegraphics[width=0.5\textwidth]{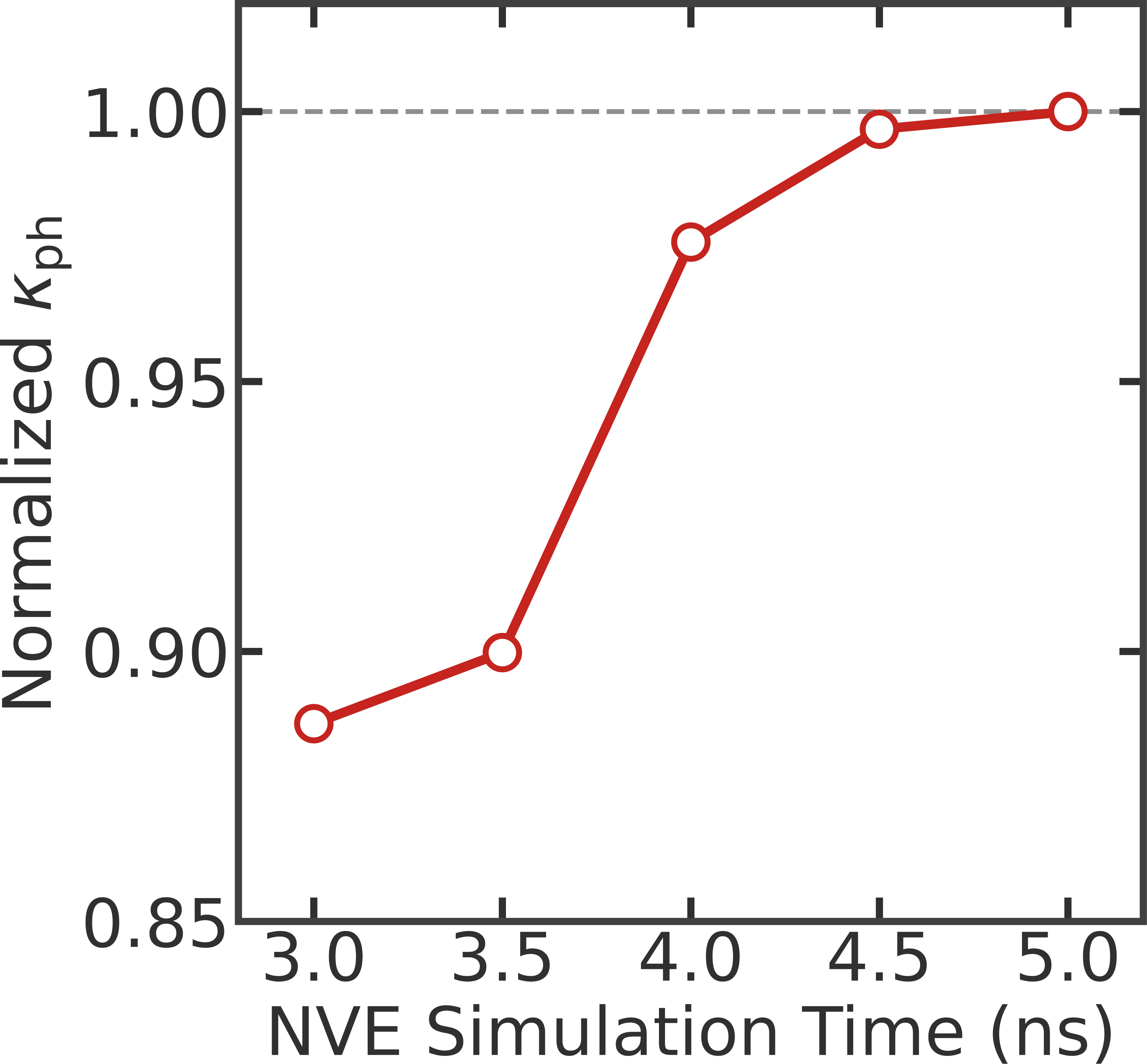}
\caption{Convergence of the calculated lattice thermal conductivity with respect to the NVE production trajectory length. The thermal conductivity is normalized by the value obtained using the 5 ns trajectory. The calculated value approaches a stable value with increasing trajectory length, with a difference of less than 1\% between the 4.5 and 5 ns trajectories.}
\label{fig:nve_convergence}
\end{figure}

To examine the convergence with respect to the NVE production trajectory length, we calculated the lattice thermal conductivity using trajectories ranging from 3 to 5 ns.
As shown in Fig.~\ref{fig:nve_convergence}, the calculated thermal conductivity gradually approaches the value obtained from the 5 ns trajectory as the trajectory length increases.
The thermal conductivity obtained using 4.5 ns differs from the 5 ns value by less than 1\%, indicating that the 5 ns NVE production trajectory used in the present calculations provides sufficient statistical sampling.

\begin{figure}[!htb]
\centering
\includegraphics[width=0.75\textwidth]{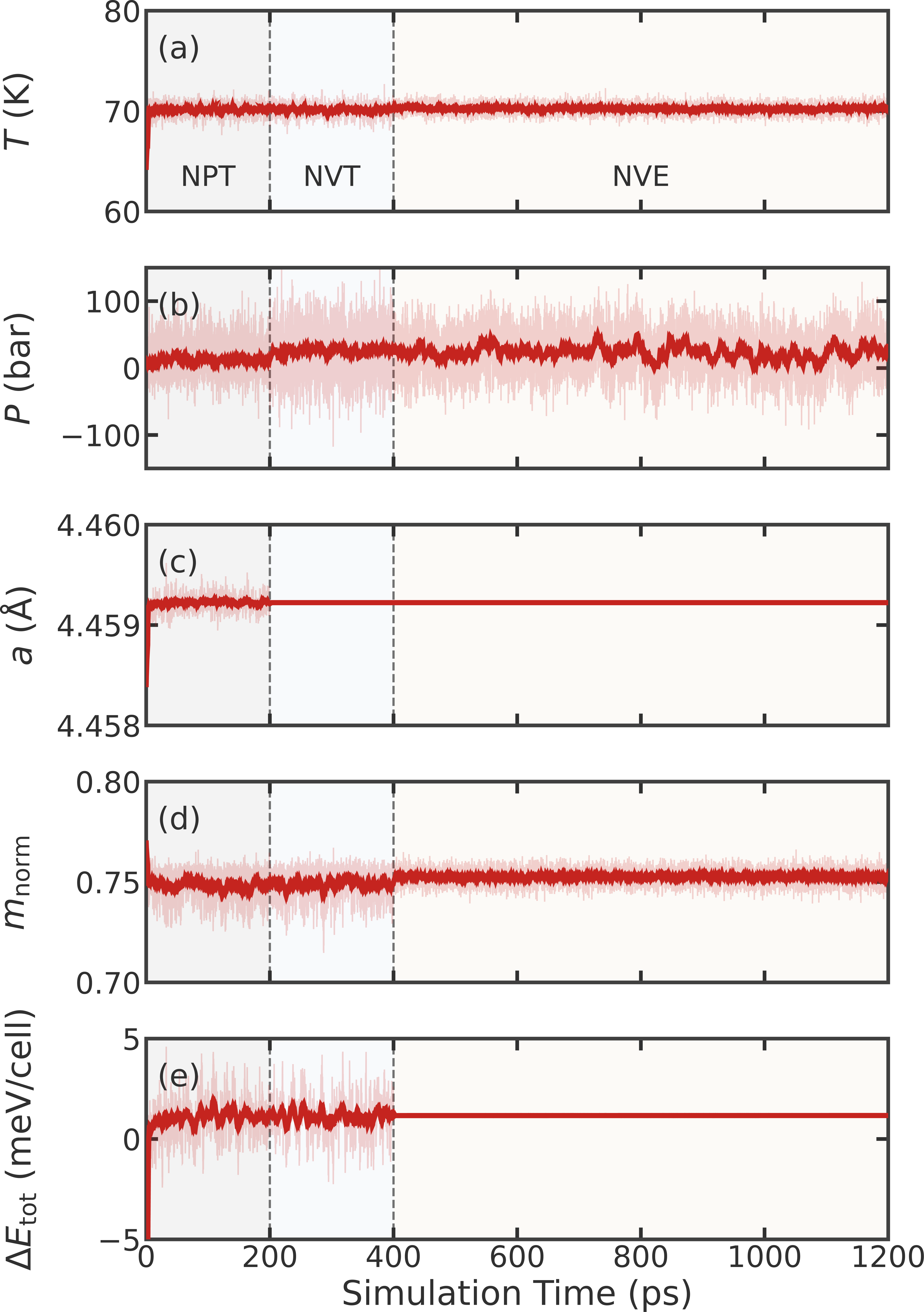}
\caption{Equilibration and stability of the trajectory used for the SED analysis. The system was equilibrated for 200 ps in the NPT ensemble and 200 ps in the NVT ensemble, followed by an 800 ps NVE production run. The time evolution of (a) temperature, (b) pressure, (c) lattice parameter, (d) normalized magnetic order parameter, and (e) change in the total energy per simulation cell is shown. The dashed vertical lines separate the NPT, NVT, and NVE stages. The thermodynamic, structural, and magnetic quantities remain stable during the production trajectory, while the total energy exhibits no noticeable drift in the NVE ensemble.}
\label{fig:sed_convergence}
\end{figure}

To verify the equilibration procedure and the stability of the trajectory used for the SED analysis, we further examined the evolution of thermodynamic, structural, and magnetic quantities throughout the simulation.
The system was first equilibrated for 200 ps in the NPT ensemble, followed by 200 ps in the NVT ensemble, and an additional 800 ps NVE trajectory was then generated for the SED analysis.
As shown in Fig.~\ref{fig:sed_convergence}, the temperature, pressure, lattice parameter, and normalized magnetic order parameter reach stable values during equilibration and remain stable during the subsequent NVE production run.
The total energy also exhibits no noticeable drift in the NVE ensemble, confirming that the equilibrated configuration and the 800 ps NVE trajectory are suitable for the SED calculations.

\section{Magnetic Transition Temperature Calculations}
\label{app:transition}
The magnetic transition temperatures were determined from the temperature dependence of the magnetic order parameters obtained from SLD and atomistic spin dynamics (ASD) simulations~\cite{evans2014atomistic}.
For the antiferromagnetic NiO and MnO systems, the normalized staggered magnetization $L(T)/L(0)$ was used, whereas the normalized magnetization $M(T)/M(0)$ was used for ferromagnetic EuO.
The calculated order parameters were interpolated as continuous functions of temperature using cubic splines, and the transition temperature was identified from the temperature at which the magnitude of the corresponding temperature derivative reaches its maximum,
\begin{equation}
T_{\mathrm{N}}
=
\underset{T}{\operatorname{arg\,max}}
\left|
\frac{dL(T)}{dT}
\right|
\end{equation}
for NiO and MnO, and
\begin{equation}
T_{\mathrm{C}}
=
\underset{T}{\operatorname{arg\,max}}
\left|
\frac{dM(T)}{dT}
\right|
\end{equation}
for EuO.

\begin{figure}[!htb]
\centering
\includegraphics[width=\textwidth]{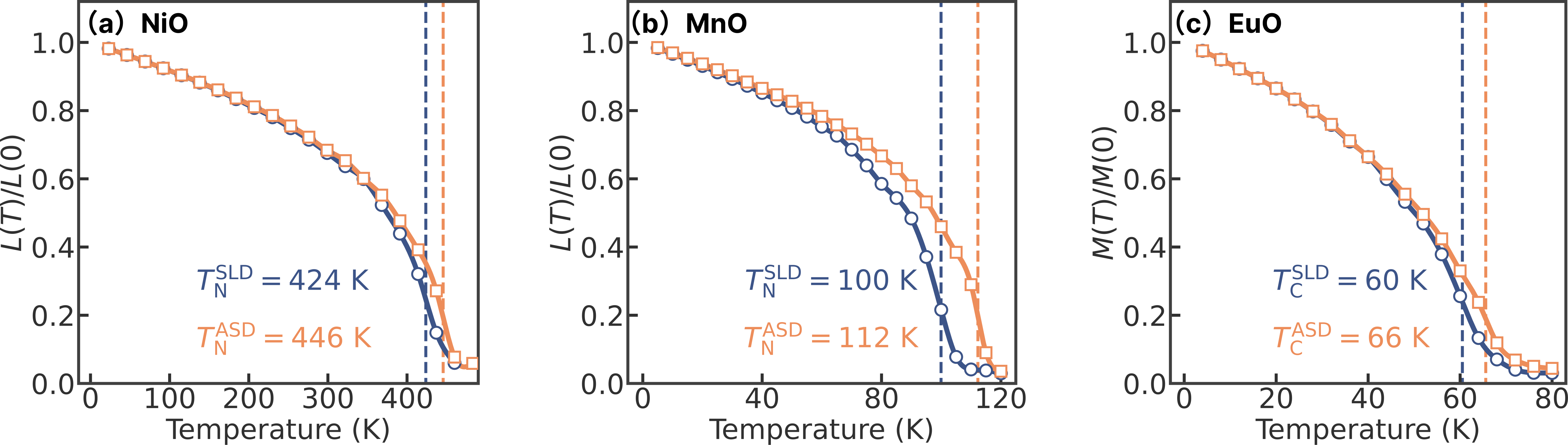}
\caption{(a,b) Normalized staggered magnetization, $L(T)/L(0)$, for antiferromagnetic NiO and MnO, respectively. (c) Normalized magnetization, $M(T)/M(0)$, for ferromagnetic EuO. Open symbols show the simulation results and solid curves show the cubic-spline interpolations. Blue and orange curves correspond to SLD and ASD simulations, respectively. The transition temperatures are determined from the maximum magnitude of the temperature derivative and are indicated by the corresponding dashed lines. The SLD transition temperatures are $T_{\mathrm{N}}=424$ K for NiO, $T_{\mathrm{N}}=100$ K for MnO, and $T_{\mathrm{C}}=60$ K for EuO, while the corresponding ASD values are 446 K, 112 K, and 66 K.}
\label{fig:transition_temperature}
\end{figure}

The resulting SLD transition temperatures are $T_{\mathrm{N}}=424$ K for NiO, $T_{\mathrm{N}}=100$ K for MnO, and $T_{\mathrm{C}}=60$ K for EuO, while the corresponding ASD values are 446 K, 112 K, and 66 K, respectively, as shown in Fig.~\ref{fig:transition_temperature}.

\section{Spin Thermal Conductivity}
\label{app:spinkappa}

\begin{figure}[!htb]
\centering
\includegraphics[width=0.75\textwidth]{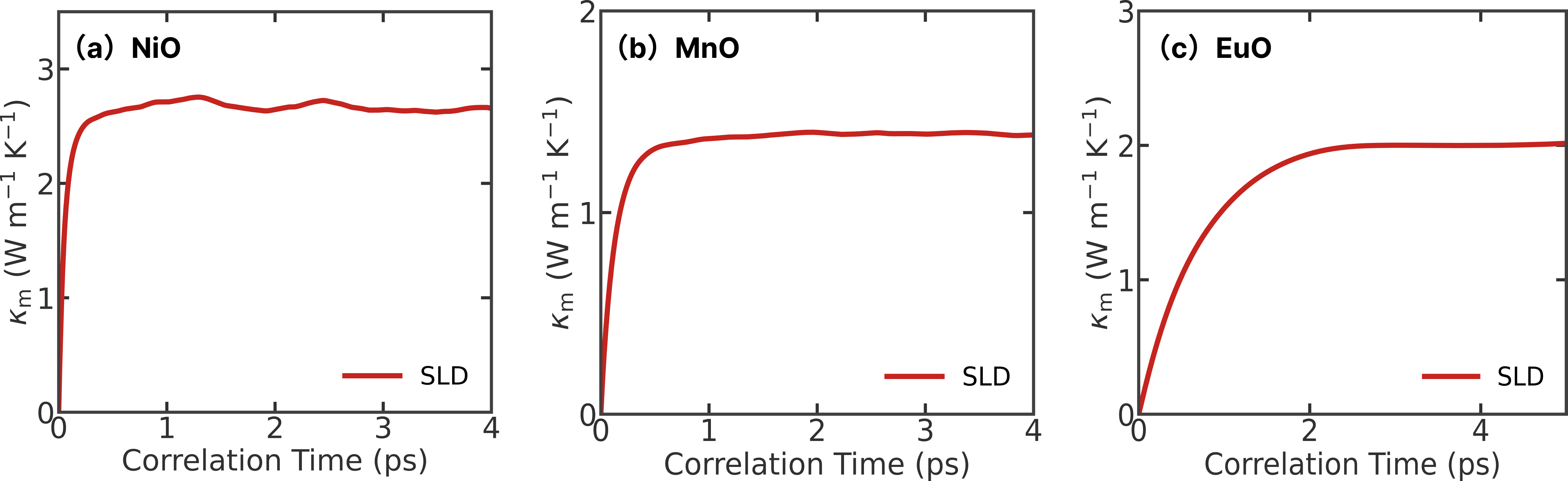}
\caption{
Cumulative spin thermal conductivity, $\kappa_{\mathrm{m}}$, obtained from SLD simulations as a function of correlation time for (a) NiO at 300~K, (b) MnO at 70~K, and (c) EuO at 40~K.
The spin thermal conductivity reaches converged values of approximately 2.7, 1.4, and 2.0~W\,m$^{-1}$\,K$^{-1}$ for NiO, MnO, and EuO, respectively.
}
\label{fig:spinkappa}
\end{figure}

To assess the contribution of spin excitations to thermal transport, we calculated the spin thermal conductivity from the SLD trajectories.
Figure~\ref{fig:spinkappa} shows the cumulative spin thermal conductivity, $\kappa_{\mathrm{m}}$, as a function of correlation time for NiO at 300~K, MnO at 70~K, and EuO at 40~K.
Compared with the lattice thermal conductivity, the cumulative spin thermal conductivity increases much more rapidly with correlation time for all three materials and quickly reaches a stable plateau.
The converged values of $\kappa_{\mathrm{m}}$ are approximately 2.7, 1.4, and 2.0~W\,m$^{-1}$\,K$^{-1}$ for NiO, MnO, and EuO, respectively.
These spin thermal conductivities are substantially smaller than the corresponding lattice thermal conductivities at the same temperatures discussed in the main text.
The spin contribution therefore represents only a minor fraction of the total thermal transport under the conditions considered here.
Accordingly, the discussion in the main text focuses primarily on the lattice thermal conductivity and its modification by spin--lattice coupling.

\section{Magnon-Phonon Scattering Pathways in Antiferromagnetic and Ferromagnetic Systems
}
\label{app:scattering_pathways}
For simplicity, we consider only the lowest-order three-particle scattering processes.
Accordingly, the magnon-phonon scattering pathways can be classified into magnon number conserving and magnon number nonconserving processes~\cite{streib2019magnon}.
The magnon number conserving process corresponds to ($m+ph\leftrightarrow m'$), where one magnon scatters with one phonon to generate another magnon. 
The magnon number nonconserving process corresponds to ($m+m'\leftrightarrow ph$), where two magnons scatter into one phonon, together with the reverse process.
Based on angular momentum conservation in scattering processes, magnon number nonconserving processes are absent in ferromagnetic systems when only exchange-mediated spin-phonon coupling is considered.
Here, we theoretically compare exchange-mediated spin-phonon coupling in ferromagnetic and antiferromagnetic systems, for which the leading-order coupling Hamiltonian is expressed as
\begin{equation}
    \mathcal{H}_{\mathrm{sp}}
    \sim
    -
    \sum_{i<j}
    \frac{\partial J(r_{ij})}{\partial r_{ij}}
    \delta r_{ij}
    \bm{s}_i\cdot\bm{s}_j .
\end{equation}
The spin product can be rewritten  as \begin{equation} \bm{s}_i\cdot\bm{s}_j = s_i^z s_j^z + \frac{1}{2} \left( s_i^+ s_j^- + s_i^- s_j^+ \right). \label{eq:sisj}\end{equation} 
where the spin raising and lowering operators are expressed using the Holstein--Primakoff transformation~\cite{holstein1940field} as
\begin{equation}
    s_i^+ = \sqrt{2s}\,a_i, 
    \qquad
    s_i^- = \sqrt{2s}\,a_i^{\dagger},
    \qquad
    s_i^z = s - a_i^{\dagger}a_i .
    \label{eq:H-P}
\end{equation}
Here, $a_i^{\dagger}$ is the magnon creation operator and $a_i$ is the magnon annihilation operator.
For ferromagnetic systems, substituting Eq.~\ref{eq:H-P} into Eq.~\ref{eq:sisj} gives
\begin{equation}
    \bm{s}_i\cdot\bm{s}_j
    =
    \left(s-a_i^{\dagger}a_i\right)
    \left(s-a_j^{\dagger}a_j\right)
    +
    s\left(a_i^{\dagger}a_j+a_j^{\dagger}a_i\right).
\end{equation}
For antiferromagnetic systems, the two sublattices have opposite equilibrium spin directions.
Considering a spin pair with site $i$ on the up sublattice and site $j$ on the down sublattice, the Holstein--Primakoff transformation~\cite{holstein1940field} can be written as 
\begin{equation} s_i^+ = \sqrt{2s}\,a_i, \qquad s_i^- = \sqrt{2s}\,a_i^{\dagger}, \qquad s_i^z = s-a_i^{\dagger}a_i , \end{equation} \begin{equation} s_j^+ = \sqrt{2s}\,c_j^{\dagger}, \qquad s_j^- = \sqrt{2s}\,c_j, \qquad s_j^z = -s+c_j^{\dagger}c_j , \end{equation} 
Substituting these expressions into Eq.~\ref{eq:sisj} gives \begin{equation} \bm{s}_i\cdot\bm{s}_j = \left(s-a_i^{\dagger}a_i\right) \left(-s+c_j^{\dagger}c_j\right) + s\left(a_i c_j+a_i^{\dagger}c_j^{\dagger}\right). \end{equation}

The pair-distance fluctuation can be approximated as the projection of the relative atomic displacement along the bond direction:
\begin{equation}
    \delta r_{ij}
    =
    \hat{\mathbf{r}}_{ij}\cdot
    \left(
    \mathbf{u}_i-\mathbf{u}_j
    \right).
\end{equation}
Since the atomic displacement can be approximately represented by phonon creation and annihilation operators as $\mathbf{u}\sim b+b^{\dagger}$, the pair-distance fluctuation can also be written as $\delta r_{ij}\sim b+b^{\dagger}$.

Therefore, considering the lowest-order three-particle scattering processes, the coupling Hamiltonians for ferromagnetic and antiferromagnetic systems can be written schematically as
\begin{equation}
    \mathcal{H}_{\mathrm{sp,FM}}
    \sim
    \left( b+b^{\dagger} \right)
    \left( a^{\dagger}a \right),
\end{equation}
and
\begin{equation}
    \mathcal{H}_{\mathrm{sp,AFM}}
    \sim
    \left( b+b^{\dagger} \right)
    \left(
    a^{\dagger}a
    +
    c^{\dagger}c
    +
    ac
    +
    a^{\dagger}c^{\dagger}
    \right),
    \label{sp,AFM}
\end{equation}
respectively.
In other words, exchange-mediated spin-phonon coupling in ferromagnetic systems only gives rise to magnon number conserving scattering processes, while both magnon number conserving and magnon number nonconserving scattering processes can occur in antiferromagnetic systems.

\section{Phonon--magnon scattering phase space}
\label{app:phase_space}

Here we derive the momentum- and energy-conservation conditions for the phonon--magnon scattering processes introduced in the main text.
For the magnon-number-conserving process in the FM state, crystal-momentum conservation requires
\begin{equation}
\mathbf{k}' = \mathbf{k}+\mathbf{q}-\mathbf{G},
\end{equation}
where $\mathbf{G}$ is a reciprocal lattice vector.
The cases $\mathbf{G}=0$ and $\mathbf{G}\neq 0$ correspond to normal and Umklapp processes, respectively.
Energy conservation further requires
\begin{equation}
\omega_{\mathrm{ph}}(\mathbf{q})
=
\left|
\omega_{\mathrm{m}}^{\mathrm{FM}}(\mathbf{k}')
-
\omega_{\mathrm{m}}^{\mathrm{FM}}(\mathbf{k})
\right|.
\end{equation}
Because the magnon dispersion is periodic in reciprocal space,
$\omega_{\mathrm{m}}^{\mathrm{FM}}(\mathbf{k}+\mathbf{G})
=
\omega_{\mathrm{m}}^{\mathrm{FM}}(\mathbf{k})$,
the magnon energy transfer can equivalently be written as
\begin{equation}
\Delta\omega_{\mathrm{m}}^{\mathrm{FM}}
=
\omega_{\mathrm{m}}^{\mathrm{FM}}(\mathbf{k}+\mathbf{q})
-
\omega_{\mathrm{m}}^{\mathrm{FM}}(\mathbf{k}).
\end{equation}
To characterize the relative energy scales of the magnetic and lattice excitations, we define
\begin{equation}
R
=
\frac{\omega_{\mathrm{m,max}}}
{\omega_{\mathrm{ph,max}}}.
\end{equation}
Using the FM dispersion defined in Eq.~\ref{eq:FM},
\begin{equation}
\Delta\omega_{\mathrm{m}}^{\mathrm{FM}}
=
\frac{\omega_{\mathrm{m,max}}}{3}
\sum_{\alpha=x,y,z}
\sin\left(k_\alpha+\frac{q_\alpha}{2}\right)
\sin\left(\frac{q_\alpha}{2}\right).
\end{equation}
Applying the Cauchy--Schwarz inequality gives
\begin{equation}
\left|
\Delta\omega_{\mathrm{m}}^{\mathrm{FM}}
\right|
\le
\omega_{\mathrm{m,max}}
\left[
\frac{
\sum_{\alpha=x,y,z}\sin^2(q_\alpha/2)
}{3}
\right]^{1/2}
=
R\,\omega_{\mathrm{ph}}(\mathbf{q}).
\end{equation}
Therefore, for $R<1$ and any nonzero $\mathbf{q}$,
\begin{equation}
\left|
\Delta\omega_{\mathrm{m}}^{\mathrm{FM}}
\right|
<
\omega_{\mathrm{ph}}(\mathbf{q}),
\end{equation}
so that the energy-conservation condition cannot be satisfied.
Because this upper bound is attainable for appropriate $\mathbf{k}$ and $\mathbf{q}$, the FM magnon-number-conserving scattering channel is first allowed at
\begin{equation}
R_{\mathrm{c}}=1.
\end{equation}
For the magnon-number-conserving channel in the AFM state, crystal-momentum conservation requires
\begin{equation}
\mathbf{k}+\mathbf{q}
=
\mathbf{k}'+\mathbf{G}_{\mathrm{m}},
\end{equation}
where $\mathbf{G}_{\mathrm{m}}$ is a reciprocal lattice vector of the magnetic lattice.
The corresponding energy-conservation condition is
\begin{equation}
\omega_{\mathrm{ph}}(\mathbf{q})
=
\left|
\omega_{\mathrm{m},j}^{\mathrm{AFM}}(\mathbf{k}')
-
\omega_{\mathrm{m},i}^{\mathrm{AFM}}(\mathbf{k})
\right|.
\end{equation}
For the magnon-number-nonconserving pair-creation and annihilation channel, crystal-momentum conservation requires
\begin{equation}
\mathbf{q}
=
\mathbf{k}+\mathbf{k}'+\mathbf{G}_{\mathrm{m}},
\end{equation}
while energy conservation requires
\begin{equation}
\omega_{\mathrm{ph}}(\mathbf{q})
=
\omega_{\mathrm{m},i}^{\mathrm{AFM}}(\mathbf{k})
+
\omega_{\mathrm{m},j}^{\mathrm{AFM}}(\mathbf{k}').
\end{equation}
The total AFM scattering phase space is therefore evaluated numerically as the sum of the magnon-number-conserving and magnon-number-nonconserving contributions.
\clearpage

\section*{References}
\bibliography{references.bib}

\end{document}